\documentclass[nofootinbib]{revtex4}  

\usepackage{epsfig}
\usepackage{epstopdf}
\usepackage[dvipsnames]{xcolor}
\usepackage{graphics}
\usepackage{lscape}
\usepackage{rotfloat}
\usepackage{rotating}
\usepackage{amsmath}
\usepackage{amssymb}
\usepackage{graphicx}                                                                        
\usepackage{flexisym}
\usepackage{breqn}
\usepackage{hyperref}

\newcommand{\bdm}{\begin{dmath}}
\newcommand{\edm}{\end{dmath}}
\newcommand{\bdms}{\begin{dmath*}}
\newcommand{\edms}{\end{dmath*}}
\newcommand{\bdg}{\begin{dgroup*}}
\newcommand{\edg}{\end{dgroup*}}

\def\lsim{\mathrel{\rlap{\lower4pt\hbox{\hskip1pt$\sim$}}
    \raise1pt\hbox{$<$}}}                
\def\gsim{\mathrel{\rlap{\lower4pt\hbox{\hskip1pt$\sim$}}
    \raise1pt\hbox{$>$}}}                

\def\slashed{{/}\mskip-10.0mu}

\def\smn{{\sigma_{\mu\nu}}}
\def\circe[#1]{\overset{\circ}{#1}}

\def\openone{\leavevmode\hbox{\small1\kern-3.3pt\normalsize1}}

\def\ZS{Z_{\rm S}}
\def\ZP{Z_{\rm P}}
\def\ZV{Z_{\rm V}}
\def\ZA{Z_{\rm A}}
\def\ZT{Z_{\rm T}}

\newcommand{\be}{\begin{equation}}
\newcommand{\ee}{\end{equation}}
\newcommand{\bea}{\begin{eqnarray}} 
\newcommand{\eea}{\end{eqnarray}}

\newcommand{\qslash}{{\not{\hspace{-0.001cm}q}}}

\newcommand{\Op}{\mathcal{O}} 

\newcommand*\Red[1]{\textcolor{red}{#1}}
\newcommand*\Blue[1]{\textcolor{blue}{#1}}
\newcommand*\Cyan[1]{\textcolor{cyan}{#1}}
\newcommand*\Magenta[1]{\textcolor{magenta}{#1}}

\definecolor{mygreen}{RGB}{30,160,10}

\definecolor{myorange}{RGB}{255,130,0}
\newcommand*\MyOrange[1]{\textcolor{myorange}{#1}}

\begin{document}

\title{Singlet vs Nonsinglet Perturbative Renormalization \\
of Fermion Bilinears}

\author{M. Constantinou$^{a,b}$, M. Hadjiantonis$^{a,c}$, H. Panagopoulos$^a$ and G. Spanoudes$^a$}

\vskip 0.25cm

\email{marthac@temple.edu, mhadjian@umich.edu, haris@ucy.ac.cy, spanoudes.gregoris@ucy.ac.cy}

\vskip 0.25cm
\affiliation{
$\,$ \\
\llap{$^a$} $Department\,\, of\,\, Physics,\,\, University\,\, of\,\,Cyprus,\,\, POB\,\, 20537,\,\, 1678\,\, Nicosia,\,\, Cyprus$\\
\llap{$^b$} $Present \ address: \ Department \ of \ Physics, \ Temple \ University, \ Philadelphia, \ PA\,\, 19122 - 1801, \ USA$\\
\llap{$^c$} $Present \ address: \ Department \ of \ Physics, \ University \ of \ Michigan, \ Ann \ Arbor, \ MI \ 48109, \ USA$}


\begin{abstract}

In this paper we present the perturbative evaluation of the 
difference between the renormalization functions of flavor
singlet and nonsinglet bilinear quark operators on the lattice.
The computation is performed to two loops and to lowest order
in the lattice spacing, for a class of improved lattice actions, 
including Wilson, tree-level (TL) Symanzik and Iwasaki gluons, 
twisted mass and SLiNC Wilson fermions, as well as staggered fermions 
with twice stout-smeared links. In the staggered formalism, the stout smearing 
procedure is also applied to the definition of bilinear operators. 

\end{abstract}

\maketitle

\section{Introduction}

In this work we study the renormalization of fermion bilinears 
${\cal O}_\Gamma = \bar{\psi}\Gamma\psi$ on the lattice, where  
$\Gamma = \openone,\,\gamma_5,\,\gamma_{\mu},
\,\gamma_5\,\gamma_{\mu},\, \gamma_5\,\sigma_{\mu\,\nu} \ (\sigma_{\mu \nu} = [\gamma_\mu, \gamma_\nu]/2i)$.  
We consider flavor
singlet ($\sum_f\bar\psi_f\Gamma\psi_f$, f: flavor index) as well as
nonsinglet ($\bar\psi_{f_1} \Gamma \psi_{f_2}, f_1 \neq f_2$) operators, to two
loops in perturbation theory. More specifically, we compute the difference between the renormalization functions of singlet and nonsinglet operators.
Our calculations were performed making use of a large family of
lattice actions, including Symanzik improved gluons, Wilson and staggered
fermions with stout links, and clover fermions; twisted
mass actions (with Iwasaki or tree-level Symanzik gluons) and
the SLiNC action are members of this family. This work is a continuation to our conference paper \cite{Constantinou:2014rka}, in which we presented our perturbative results using Wilson/clover fermions.  

\bigskip

The most demanding part of this study is the computation of the
2-point Green's functions of ${\cal O}_\Gamma$\,, up to two loops. From
these Green's 
functions we extract the renormalization functions for ${\cal O}_\Gamma\,$: $Z_\Gamma^{L,Y}$ ($L$: lattice regularization, $Y$ 
$(= RI^{\prime},\ \overline{MS})$: renormalization schemes). As a
check on our results, we have computed them in an arbitrary covariant
gauge. Our expressions can be
generalized, in a straightforward manner, to fermionic fields in an
arbitrary representation.

\bigskip

Flavor singlet operators are relevant for a number of hadronic
properties including, e.g., topological features or the spin structure
of hadrons. Matrix elements of such operators are notoriously
difficult to study via numerical simulations, due to the presence of
(fermion line) disconnected diagrams, which in principle require
evaluation of the full fermion propagator. In recent years
there has been some progress in the numerical study of flavor singlet
operators; furthermore, for some of them, a nonperturbative estimate
of their renormalization has been obtained using the Feynman-Hellmann
relation~\cite{Chambers:191759}. Perturbation theory can give an important cross
check for these estimates, and provide a prototype for other
operators which are more difficult to renormalize nonperturbatively.

\bigskip

Given that for the renormalization of flavor nonsinglet operators one
can obtain quite accurate nonperturbative estimates, we will focus on the
perturbative evaluation of the \emph{difference} between the
flavor singlet and nonsinglet renormalization; this difference first
shows up at two loops.

\bigskip

Perturbative computations beyond one loop for Green's functions
with nonzero external momenta are technically quite involved, and
their complication is greatly increased when improved gluon and
fermion actions are employed. For fermion bilinear operators, the only
two-loop computations in standard perturbation theory thus far have been
performed by our group~\cite{Skouroupathis:2007jd, Skouroupathis:2008mf}, employing Wilson gluons and
Wilson/clover fermions. Similar investigations have been carried out
in the context of stochastic perturbation theory~\cite{DiRenzo:2014}.

\bigskip

Staggered fermions entail additional complications as compared to Wilson fermions. 
In particular, the fact that fermion degrees of freedom are distributed over
neighbouring lattice points requires the introduction of link variables in the
definition of gauge invariant fermion bilinears, with a corresponding increase in the
number of Feynman diagrams. In addition, the appearance of 16 (rather than 1) poles
in the fermion propagator leads to a rather intricate structure of divergent contributions 
in two-loop diagrams. 

\bigskip

A novel aspect of the calculations is that the gluon links, which appear both in the staggered fermion action and in the definition of the bilinear operators in the staggered basis, are improved by applying a stout smearing procedure up to two times, iteratively. Compared to most other improved formulations of staggered fermions, the stout smearing action leads to smaller taste violating effects \cite{Aoki:2005vt,Borsanyi:2011bm,Bazavov:2012zad}. Application of stout improvement on staggered fermions thus far has been explored, by our group, only to one-loop computations \cite{Constantinou:2013pba}; a two-loop computation had never been investigated before.  

\bigskip

Further composite fermion operators of interest, to which one can apply our
perturbative techniques, are higher dimension bilinears such as:
$\bar{\psi}\Gamma\,D^\mu\psi$ (appearing in hadron structure
functions) and four-fermion operators such as: 
$(\bar{s}\,\Gamma_1\,d)\,(\bar{s}\,\Gamma_2\,d)$ (appearing in $\Delta
S = 2$ transitions, etc.); in these cases, complications such as
operator mixing greatly hinder nonperturbative methods of
renormalization, making a perturbative approach all that more essential.

\bigskip

The outline of this paper is as follows: Section II presents a brief
theoretical background in which we introduce the formulation of the
actions and of the operators which we employ, as well as all necessary 
definitions of renormalization schemes and of the quantities to compute. 
Section III contains the calculational procedure and the results which are obtained.
In Section IV we discuss our results and we plot several graphs for certain values 
of free parameters, like the stout and clover coefficients. Finally, we
conclude with possible future extensions of our work. For completeness, 
we have included two Appendices containing: A: the formulation of the fermion action and 
the bilinear operators in the staggered basis, along with a compendium of useful relations and B: evaluation of a basis of nontrivial divergent two-loop Feynman diagrams, which appeared 
in our calculation in the staggered formalism. 

\bigskip

\section{Formulation}

\subsection{Lattice actions}

In our calculation we made use of two totally different formulations 
of the fermion action on the lattice: Wilson and staggered formulation. 
In the Wilson formulation, we used the SLiNC fermion action \cite{Horsley:2008ap}, i.e., we 
introduced stout gluon links in the standard Wilson action and also we 
added the clover term. In standard notation, it reads:  
\begin{eqnarray}
S_{\rm WF} = \hspace{-0.1cm}&-&\hspace{-0.1cm} \frac{a^3}{2} \sum_{x, \mu} \, \overline{\psi} (x) 
\Big[(r - \gamma_\mu ) \widetilde{U}_\mu (x) \ \psi (x + a \hat{\mu})
+ (r + \gamma_\mu ) \widetilde{U}_\mu^\dagger (x - a \hat{\mu}) 
\ \psi (x - a \hat{\mu}) - 2 r \ \psi (x)\Big] \nonumber \\ [2ex]
&-& \frac{a^5}{4} \sum_{x, \mu, \nu} c_{SW} \ \overline{\psi} (x) 
\ \sigma_{\mu \nu} \ \hat{G}_{\mu \nu} (x) \ \psi (x)\,
\label{WFaction1}
\end{eqnarray}
where
\be 
\hat{G}_{\mu \nu} (x) = \Big[ Q_{\mu \nu} (x) - Q_{\nu \mu} (x) \Big] / (8 a^2)
\ee
and
\begin{eqnarray}
Q_{\mu \nu} &=& U_\mu (x) \ U_\nu (x + a \hat{\mu}) \ U_\mu^\dagger (x + a \hat{\nu}) \ U_\nu^\dagger (x) \nonumber \\ [2ex] 
&+& U_\nu (x) \ U_\mu^\dagger (x + a \hat{\nu} - a \hat{\mu}) \ U_\nu^\dagger (x - a \hat{\mu}) \ U_\mu (x - a \hat{\mu}) \nonumber \\ [2ex]
&+& U_\mu^\dagger (x - a \hat{\mu}) \ U_\nu^\dagger (x - a \hat{\mu} -a \hat{\nu}) \ U_\mu (x - a \hat{\mu} - a \hat{\nu}) \ U_\nu (x - a \hat{\nu}) \nonumber \\ [2ex]
&+& U_\nu^\dagger (x - a \hat{\nu}) \ U_\mu (x - a \hat{\nu}) \ U_\nu (x + a \hat{\mu} - a \hat{\nu}) \ U_\mu^\dagger (x)
\end{eqnarray}
The fermion action may also contain standard and twisted mass terms, 
but they only contribute beyond two loops to the difference between
flavor singlet and nonsinglet renormalizations; this is true in 
mass-independent schemes, such as $RI'$ and $\overline{MS}$, in which
renormalized masses vanish. The quantities $\widetilde{U}_\mu (x)$, 
appearing above, are stout gluon links, defined as \cite{Morningstar:2003gk}:
\be
\widetilde{U}_\mu(x) = e^{i\,Q_\mu(x)}\,U_\mu(x)\,
\label{stoutlink}
\ee
where
\be
  Q_\mu(x)=\frac{\omega}{2\,i} \left[V_\mu(x) U_\mu^\dagger(x) -
  U_\mu(x)V_\mu^\dagger(x) -\frac{1}{N_c} {\rm Tr} \,\Big(V_\mu(x)
  U_\mu^\dagger(x) -  U_\mu(x)V_\mu^\dagger(x)\Big)\right] \, 
\label{Q_def}
\ee
$V_\mu(x)$ represents the sum over all staples associated with the
link $U_\mu(x)$ and $N_c$ is the number of colors. Following common practice, we henceforth set the Wilson parameter r equal to 1. Both the stout coefficient $\omega$ and the clover 
coefficient $c_{SW}$ will be treated as free parameters, for wider applicability of the results.  


\bigskip

In the staggered formulation of fermion action, we introduced ``doubly" stout gluon links 
in the naive staggered action. In standard notation, it reads:  
\be
S_{\rm SF} = a^4 \sum_{x,\mu} \frac{1}{2a} \, \overline{\chi} (x) 
\ \eta_\mu (x) \ \Big[ \widetilde{\widetilde{U}}_\mu (x) \ \chi (x + a \hat{\mu})
- \widetilde{\widetilde{U}}_\mu^\dagger (x - a \hat{\mu}) \ \chi (x - a \hat{\mu}) \Big] 
+ a^4 \sum_{x} m \ \overline{\chi} (x) \ \chi (x)\,
\label{SFactionimpr1}
\ee
where $\chi (x)$ is a one-component fermion field, and $\eta_\mu (x) = (-1)^{\sum_{\nu < \mu} n_\nu}$ 
[$ x =(a\,n_1,a\,n_2,a\,n_3,a\,n_4),\quad n_i\,\, \epsilon\,\, {\mathbb Z}\,$]. 
Just as in the case of Wilson fermions, the mass term will be irrelevant, since we will apply
mass-independent renormalization schemes.
In Appendix A we remind the reader of the relation between the staggered field $\chi(x)$ 
and the standard fermion field $\psi (x)$. The gluon links 
$\widetilde{\widetilde{U}}_\mu (x)$, appearing above, 
are doubly stout links, defined as:
\be
\widetilde{\widetilde{U}}_\mu(x) = e^{i\,\widetilde{Q}_\mu(x)}\,\widetilde{U}_\mu(x)\,
\ee
where $\widetilde{U}_\mu(x)$ is the singly stout link, defined in Eq. \eqref{stoutlink}
and $\widetilde{Q}_\mu(x)$ is defined as in Eq.\eqref{Q_def}, but using 
$\widetilde{U}_\mu$ as links (also in the construction of $V_\mu$). To obtain results 
that are as general as possible, we use different stout parameters, $\omega$, in the first 
($\omega_1$) and the second ($\omega_2$) smearing iteration. 


\bigskip

For gluons, we employ a Symanzik improved action, of the form \cite{Horsley:2004mx}:
\bea
\hspace{-1cm}
S_G=\frac{2}{g_0^2} \Bigl[ &c_0& \sum_{\rm plaq.} {\rm Re\,Tr\,}\{1-U_{\rm plaq.}\}
\,+\, c_1 \sum_{\rm rect.} {\rm Re \, Tr\,}\{1- U_{\rm rect.}\} 
\nonumber \\ 
+ &c_2& \sum_{\rm chair} {\rm Re\, Tr\,}\{1-U_{\rm chair}\} 
\,+\, c_3 \sum_{\rm paral.} {\rm Re \,Tr\,}\{1-U_{\rm paral.}\}\Bigr]\,
\label{Symanzik}
\eea
\begin{figure}[!htbp]
\centering
\includegraphics[scale=1.3]{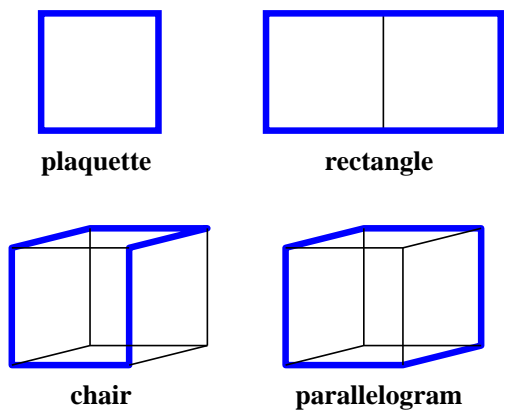}
\caption{The 4 Wilson loops of the gluon action.}
\label{figSym}
\end{figure} 
where $U_{\rm plaq.}$ is the 4-link Wilson loop and $U_{\rm rect.}$, $U_{\rm chair}$, 
$U_{\rm paral.}$ are the three possible independent 6-link Wilson loops (see Fig. ~\ref{figSym}).
The Symanzik coefficients $c_i$ satisfy the following normalization condition:
\be
c_0 + 8 c_1 + 16 c_2 + 8 c_3 = 1\,
\label{norm}
\ee

The algebraic part of our computation was carried out for generic values of $c_i$; for the numerical integration over loop momenta we selected a number of commonly used sets of values, some of which are shown in Table \ref{tab1}. We have also performed our calculations for some other values of $c_i$, relevant to the TILW (tadpole improved L\"uscher - Weisz) family of actions.

\begin{table}[h]
\begin{center}
\begin{minipage}{13cm}
\begin{tabular}{lr@{}lr@{}lr@{}lr@{}l}
\hline
\hline
\multicolumn{1}{c}{Gluon action}&
\multicolumn{2}{c}{$\qquad c_0$}&
\multicolumn{2}{c}{$\quad c_1$}&
\multicolumn{2}{c}{$\quad c_2$}&
\multicolumn{2}{c}{$\quad c_3$}\\
\hline
\hline
\multicolumn{1}{c}{Wilson}&
\multicolumn{2}{c}{$\qquad 1$}&
\multicolumn{2}{c}{$\quad 0$}&
\multicolumn{2}{c}{$\quad 0$}&
\multicolumn{2}{c}{$\quad 0$}\\
\multicolumn{1}{c}{TL Symanzik}&
\multicolumn{2}{c}{$\qquad 5/3$}&
\multicolumn{2}{c}{$\quad -1/12$}&
\multicolumn{2}{c}{$\quad 0$}&
\multicolumn{2}{c}{$\quad 0$}\\
\multicolumn{1}{c}{Iwasaki}&
\multicolumn{2}{c}{$\qquad 3.648$}&
\multicolumn{2}{c}{$\quad -0.331$}&
\multicolumn{2}{c}{$\quad 0$}&
\multicolumn{2}{c}{$\quad 0$}\\
\multicolumn{1}{c}{DBW2}&
\multicolumn{2}{c}{$\qquad 12.2688$}&
\multicolumn{2}{c}{$\quad -1.4086$}&
\multicolumn{2}{c}{$\quad 0$}&
\multicolumn{2}{c}{$\quad 0$}\\
\hline
\hline
\end{tabular}
\end{minipage}
\end{center}
\caption{Selected sets of values for Symanzik coefficients}
\label{tab1}
\end{table}  

\subsection{Definition of bilinear operators in the staggered basis}

In the staggered formalism a physical fermion field $\psi (x)$ has 
``taste" components, besides Dirac components (see Appendix A). Hence, the fermion 
bilinear operators, written in terms of fermion fields with taste
$\psi_{\alpha,t} (x)$ ($\alpha$: Dirac index, $t$: taste index), have the 
following form:
\be
{\cal O}_{\Gamma,\xi}=\bar \psi (x)\,\left(\Gamma\otimes\xi\right)\,\psi (x)\,
\label{O_G}
\ee
where $\Gamma$ and $\xi$ are arbitrary $4\times4$ matrices acting on
the Dirac and taste indices of $\psi_{\alpha,t} (x)$, respectively. After
transforming to the staggered basis, the operator ${\cal O}_{\Gamma,\xi}$ 
can be written as \cite{Patel:1992vu}:
\bea
{\cal O}_{\Gamma,\xi} &=& \sum_{C,D}
\bar\chi(y)_C\,\left(\overline{\Gamma
  \otimes\xi}\right)_{CD}\,U_{C,D}\,\chi(y)_D\,,
\label{O_general}\\
&&\left(\overline{\Gamma\otimes\xi}\right)_{CD} \equiv  
\frac{1}{4}\,{\rm
  Tr}\left[\gamma^\dagger_C\,\Gamma\,\gamma_D\,\xi\right]\,
\label{gamma1}
\eea
where $\chi(y)_C\equiv\chi(y+aC)/4$ [$y$ denotes the position of a hypercube inside the lattice ($y_\mu\,\in\,\,2{\mathbb Z}$), C denotes the position of a fermion field component within a specific hypercube ($C_\mu \in \{0,1\}$)] and $\gamma_C = \gamma_1^{C_1}\,\gamma_2^{C_2}\,\gamma_3^{C_3}\,\gamma_4^{C_4}$. In order to ensure the gauge invariance of the above operators, one inserts the quantity $U_{C,D}$, which is the average of products of gauge link variables along all possible shortest paths connecting the sites $y+C$ and $y+D$. In this work we focus on taste-singlet operators, thus $\xi=\openone$. Explicitly, the taste-singlet bilinear operators can be written as:
\bea
\label{OS2}
{\cal O}_S(y) &=& \sum_D \bar\chi(y)_{D} \, \chi(y)_D\, \\
{\cal O}_V(y) &=& \sum_D \bar\chi(y)_{D+_{_2}\hat{\mu}}\,U_{D+_{_2}\hat{\mu},D}\,\chi(y)_D\,\eta_\mu(D)\,\\
{\cal O}_T(y) &=& \frac{1}{i} \sum_D\,\bar\chi(y)_{D+_{_2}\hat{\mu}+_{_2}\hat{\nu}}\,U_{D+_{_2}\hat{\mu}+_{_2}\hat{\nu},D} \,\chi(y)_D \,\eta_\nu(D) \,\eta_{\mu}(D+_{_2}\hat{\nu})\,,\quad \mu \neq \nu \\ \label{OT2}
{\cal O}_A(y) &=& \sum_D\bar\chi(y)_{D+_{_2}\hat{\mu}+_{_2}(1,1,1,1)}\,U_{D+_{_2}\hat{\mu}+_{_2}(1,1,1,1),D}\,\chi(y)_D\,\eta_\mu(D)\, \cdot \nonumber \\
& & \qquad \eta_1(D+_{_2}\hat{\mu})\,\eta_2(D+_{_2}\hat{\mu})\,\eta_3(D+_{_2}\hat{\mu})\,\eta_4(D+_{_2}\hat{\mu})\,\\
{\cal O}_P(y) &=& \sum_D \bar\chi(y)_{D+_{_2}(1,1,1,1)}\,U_{D+_{_2}(1,1,1,1),D}\,\chi(y)_D\,\eta_1(D)\,\eta_2(D)\,\eta_3(D)\,\eta_4(D)\, \label{OP2}
\eea
where $S (\rm Scalar),\, P (\rm Pseudoscalar),\, V (\rm Vector),\, A (\rm Axial \ Vector),\, T (\rm Tensor)$ correspond to: $\Gamma = \openone, \gamma_5, \gamma_\mu, \gamma_5 \gamma_\mu, \gamma_5 \ \sigma_{\mu \nu}$ and $a +_{_2} b \equiv (a+b)$ mod $2$. Further details on the formulation of these operators are provided in Appendix A. With the exception of the Scalar operator, the remaining operators
contain averages of products of up to 4 gluon links (in orthogonal
directions) between the fermion and the antifermion fields. Just as in the staggered fermion action, the gluon links used in the operators, are doubly stout links. We have kept the stout parameters of the action (${\omega}_{A_1},{\omega}_{A_2}$) distinct from the stout parameters of the operators (${\omega}_{O_1},{\omega}_{O_2}$), for wider applicability of the results. The presence of gluon links in the definition of bilinear operators creates new Feynman diagrams which do not appear in the Wilson formulation, leading to nontrivial contributions in our two-loop calculation.    

\subsection{Renormalization of fermion bilinear operators}

The renormalization functions $Z_\Gamma$ for lattice fermion bilinear 
operators, relate the bare operators 
$\mathcal{O}_{\Gamma_\circ} = \bar{\psi} \Gamma \psi$ to their 
corresponding renormalized continuum operators 
$\mathcal{O}_\Gamma$ via:
\be
 \mathcal{O}_\Gamma = Z_\Gamma \,\mathcal{O}_{\Gamma_\circ}
\ee
Renormalization functions of such lattice operators are necessary
ingredients in the prediction of physical probability amplitudes from 
lattice matrix elements. In order to calculate the renormalization 
functions $Z_\Gamma$, it is essential to compute the 2-point amputated Green's functions of the operators $\mathcal{O}_{\Gamma_\circ}$; they can be written in the following form:
\bea
\Sigma_{\rm S} (a q) &=& \openone \ \Sigma^{(1)}_{\rm S} (a q) \label{GreensFunctionS}\\
\Sigma_{\rm P} (a q) &=& \gamma_5 \ \Sigma^{(1)}_{\rm P} (a q) \label{GreensFunctionP}\\
\Sigma_{\rm V} (a q) &=& \gamma_{\mu} \ \Sigma^{(1)}_{\rm V} (a q) +  \frac{q^{\mu} \slashed{q}}{q^2} \ \Sigma^{(2)}_{\rm V} (a q) \\
\Sigma_{\rm A} (a q) &=& \gamma_5 \ \gamma_{\mu} \ \Sigma^{(1)}_{\rm A} (a q) + \gamma_5 \ \frac{q^{\mu} \slashed{q}}{q^2} \ \Sigma^{(2)}_{\rm  ÁA} (a q) \label{GreensFunctionA}\\
\Sigma_{\rm ÔT} (a q) &=& \gamma_5 \ \sigma_{\mu \nu} \ \Sigma^{(1)}_{\rm T} (a q) + \gamma_5 \ \frac{\slashed{q}}{q^2} (\gamma_{\mu} q_{\nu} - \gamma_{\nu} q_{\mu}) \ \Sigma^{(2)}_{\rm T} (a  q) \label{GreensFunctionT}
\eea
where $\Sigma^{(1)}_{\Gamma} = 1 + \mathcal{O}(g_{\circ}^2), \ \Sigma^{(2)}_\Gamma= {\cal O}(g_\circ^2)$, $g_\circ$: bare coupling constant. 

\bigskip

The $RI'$ renormalization scheme is defined by imposing renormalization
conditions on matrix elements at a scale $\bar{\mu}$. The 
renormalization condition giving $Z_\Gamma^{L,RI'}$ (L: Lattice) is:
\be
\lim_{a \rightarrow 0} \bigg[ Z_{\psi}^{L,RI'} \ Z_{\Gamma}^{L,RI'} \ \Sigma_{\Gamma}^{(1)} (a q) \bigg]_{\begin {smallmatrix}
\ q^2 = {\bar{\mu}}^2, \\
m = 0
\end{smallmatrix}}= 1
\label{ZGamma}
\ee
where $Z_\psi$ is the renormalization function for the fermion field ($\psi = Z_\psi^{-1 / 2} \ \psi_\circ$, $\psi (\psi_\circ)$: renormalized (bare) fermion field). Such a condition guarantees that the renormalized Green's function of ${\cal{O}}_\Gamma$ (the quantity in brackets in Eq. \ref{ZGamma}) will be a finite function of the renormalized coupling constant $g$ for all values of the momenta ($g = \mu^{- \epsilon} Z_g^{-1} g_\circ$ where $\mu$ is the mass scale introduced to ensure that $g_\circ$ has the correct dimensionality in $d=4-2\epsilon$ dimensions). Comparison between the $RI'$ and the $\overline{\rm MS}$ schemes is normally performed at the same scale $\bar{\mu} = \mu (4 \pi / e^{\gamma_E})^{1/2}$. 

\bigskip

The $RI'$ renormalization prescription, as given above, does not involve 
$\Sigma_\Gamma^{(2)}$; nevertheless, renormalizability of the theory implies that $Z_\Gamma^{L,RI'}$ will render the entire Green's function finite. An alternative prescription, more appropriate for nonperturbative renormalization, is:
\be
\lim_{a \rightarrow 0} \bigg[ Z_{\psi}^{L,RI'} Z_{\Gamma}^{L,RI'(\text{alter})} \ \frac{\text{tr} \big( \Gamma \Sigma_{\Gamma} (a q) \big)}{\text{tr} \big( \Gamma \Gamma \big)} \bigg]_{\begin {smallmatrix}
\ q^2 = {\bar{\mu}}^2, \\
m = 0
\end{smallmatrix}}= 1
\ee
where a summation over repeated indices $\mu$ and $\nu$ is understood. This scheme has the advantage of taking into account the whole bare Green's function and therefore is useful for numerical simulations where the arithmetic data for $\Sigma_\Gamma$ cannot be separated into two different structures. The two prescriptions differ between themselves (for V, A, T) by a finite amount which can be deduced from lower loop calculations combined with continuum results. 

\bigskip

Conversion of renormalization functions from $RI'$ to the $\overline{MS}$ scheme is facilitated by the
fact that renormalized Green's functions are regularization independent; thus the finite conversion
factors:
\be
C_\Gamma (g,\alpha) \equiv \frac{Z_\Gamma^{L,RI'}}{Z_\Gamma^{L,\overline{MS}}} = \frac{Z_\Gamma^{DR,RI'}}{Z_\Gamma^{DR,\overline{MS}}}
\ee
(DR: Dimensional Regularization, $\alpha$: gauge parameter) can be evaluated in DR, leading to $Z_\Gamma^{L,\overline{MS}} = Z_\Gamma^{L,RI'} / C_\Gamma (g,\alpha)$. For the Pseudoscalar and Axial Vector operators, in order to satisfy Ward identities, additional finite factors $Z_5^{\rm P} (g)$ and $Z_5^{\rm A} (g)$, calculable in DR, are required:
\be
 Z_{\rm P}^{L,\overline{MS}} = \frac{Z_{\rm P}^{L,RI'}}{C_{\rm S} Z_5^{\rm P}}, \, \quad Z_{\rm A}^{L,\overline{MS}} = \frac{Z_{\rm A}^{L,RI'}}{C_{\rm V} Z_5^{\rm A}} \label{ZPA}
\ee 
These factors are gauge independent; we also note that the value of $Z_5^A$ for the flavor singlet operator differs from that of the nonsinglet one. The values of $Z_5^P$, $Z_5^{A(\rm singlet)}$ and $Z_5^{A(\rm nonsinglet)}$, calculated in Ref. \cite{Larin:1993tq}, are:
\bea
Z_5^P (g) &=& 1 - \frac{g^2}{(4 \pi)^2} (8 c_F) + \frac{g^4}{(4 \pi)^4} \Big( \frac{2}{9} c_F N_c + \frac{4}{9} c_F N_f \Big) + \mathcal{O} (g^6) \label{Z5P}\\
Z_5^{A(\rm singlet)} (g) &=& 1 - \frac{g^2}{(4 \pi)^2} (4 c_F) + \frac{g^4}{(4 \pi)^4} \Big( 22 c_F^2 - \frac{107}{9} c_F N_c + \frac{31}{18} c_F N_f \Big) + \mathcal{O} (g^6) \label{Z5As}\\
Z_5^{A(\rm nonsinglet)} (g) &=& 1 - \frac{g^2}{(4 \pi)^2} (4 c_F) + \frac{g^4}{(4 \pi)^4} \Big( 22 c_F^2 - \frac{107}{9} c_F N_c + \frac{2}{9} c_F N_f \Big) + \mathcal{O} (g^6) \label{Z5Ans}
\eea
where $c_F \equiv (N_c^2 - 1)/(2 N_c)$ and $N_f$ is the number of flavors.

\bigskip

\section{Computation and Results}

In the previous Section, the calculation setup was presented in rather general terms.
Here we focus on the two-loop difference between flavor singlet and nonsinglet operator 
renormalization. Given that this difference first arises at two loops, we only need the
tree-level values of $Z_\psi$, $Z_g$ and of the conversion factors $C_\Gamma$, $Z_5^P$ and $Z_5^A$
($Z_\psi = Z_g = C_\Gamma = Z_5^P = Z_5^A = 1$). Since $C_\Gamma = 1$, the difference up to two loops will not depend on the renormalization scheme. In addition, for our computations we will use 
mass-independent renormalization schemes; this means that the renormalized mass of quarks
will be taken to be zero. For Wilson fermions, a zero renormalized mass entails a nonzero,
${\cal O} (g^2_\circ)$ Lagrangian mass; however, such a term does not enter the quantities
which we calculate up to ${\cal O} (g^4_\circ)$. 

\subsection{Results from the Wilson formulation of the fermion action}

\begin{figure}[htbp]
\centering
\epsfig{file=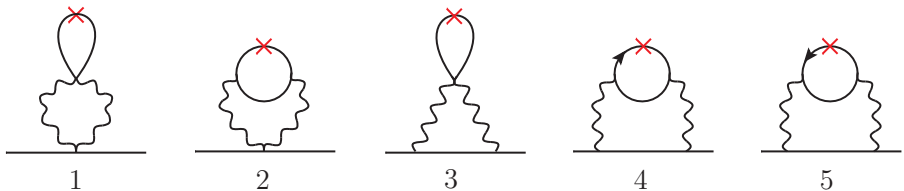,scale=0.75}
\caption{Diagrams (in Wilson formulation) contributing to the difference between flavor singlet and nonsinglet values of $Z_\Gamma$. Solid (wavy) lines represent fermions (gluons). A cross denotes insertion of the operator ${\cal{O}}_\Gamma$.}
\label{figDiagW}
\end{figure}

At first, we present the computational procedure and results for the above difference 
using the Wilson formulation of the fermion action (Eq. \ref{WFaction1}). There are 5 
two-loop Feynman diagrams contributing to this difference in the evaluation of 
the Green's functions (Eqs. \ref{GreensFunctionS} - \ref{GreensFunctionT}), shown in Fig. ~\ref{figDiagW}. They all contain an operator insertion inside a closed fermion loop, and therefore 
vanish in the flavor nonsinglet case. Note that only diagrams 4 and 5 appear in the continuum. 
In order to simplify our calculations of $Z_{\Gamma}$, we worked with $\sum_x \mathcal{O}_\Gamma (x)$ so that no momentum enters the diagrams at the operator insertion point. 

\bigskip

The above diagrams, evaluated individually, may be IR divergent, due to the presence of two-gluon propagators with the same momentum. Comparing to our previous evaluation of these diagrams
with Wilson gluons and clover fermions \cite{Skouroupathis:2007jd, Skouroupathis:2008mf}, we will 
find neither any new superficial divergences ($\ln^2(a^2 \bar{\mu}^2)$ terms) nor any new 
subdivergences ($\ln(a^2 \bar{\mu}^2)$ terms). However, the 
presence of stout links and Symanzik gluons leads to considerably longer expressions for the vertices. Also, the gluon propagator must now be inverted numerically for every choice of values for 
the Symanzik coefficients and for each value of the loop momentum 4-vector; an inversion in 
closed form exists, but it is not efficient.

\bigskip

The contribution of these diagrams to $\ZP$, $\ZV$, $\ZT$ vanishes identically just as
in continuum regularizations. Therefore, only $\ZS$ and $\ZA$ are affected. For the 
Scalar operator, our result can be written in the following form:
\bea 
\ZS^{\rm singlet} (a \bar{\mu}) &-& \ZS^{\rm nonsinglet} (a \bar{\mu}) =  \nonumber \\
&-& \frac{g_\circ^4}{(4 \pi)^4} \ c_F \ N_f \ \Bigg\lbrace(s_{00} + s_{01} \ \Red{\mathbf{ c_{SW}}} + s_{02} \ \Red{\mathbf{ c_{SW}^2}} + s_{03} \ \Red{\mathbf{ c_{SW}^3}} + s_{04} \ \Red{\mathbf{ c_{SW}^4}}) \nonumber \\
&+& (s_{10} + s_{11} \ \Red{\mathbf{ c_{SW}}} + s_{12} \ \Red{\mathbf{ c_{SW}^2}} + s_{13} \ \Red{\mathbf{ c_{SW}^3}}) \ \Blue{\boldsymbol{ \omega}} + (s_{20} + s_{21} \ \Red{\mathbf{ c_{SW}}} + s_{22} \ \Red{\mathbf{ c_{SW}^2}}) \ \Blue{\boldsymbol{ \omega^2}} \nonumber \\
&+& (s_{30} + s_{31} \ \Red{\mathbf{ c_{SW}}}) \ \Blue{\boldsymbol{ \omega^3}} + s_{40} \ \Blue{\boldsymbol{ \omega^4}} \Bigg\rbrace + \Op (g_\circ^6) 
\label{ZS}
\eea
The numerical constants $s_{ij}$ have been computed for various sets of values
of the Symanzik coefficients; their values are listed in Table \ref{tab2} for the Wilson, TL Symanzik and Iwasaki gluon actions. 
The errors quoted stem from extrapolation of the results of numerical integration over loop momenta for different lattice sizes. The extrapolation methods that we used are described in Ref. \cite{Panagopoulos:2006ky}. The computation was performed in
a general covariant gauge, confirming that the result is gauge independent, as it should be in $\overline{MS}$.
We note from Eq. \eqref{ZS} that even single logarithms are absent, and thus the result is scale independent; this is consistent with the fact that the corresponding difference for the Scalar operator in dimensional regularization is absent. 

\bigskip

\begin{table}[!ht]
\begin{center}
\begin{tabular}{|c|ccc||c|ccc|}
\hline
 & Wilson & TL Symanzik & Iwasaki & & Wilson & TL Symanzik & Iwasaki  \\
\hline
\ $s_{00}$ \quad & 107.76(2) & 76.29(1) & 42.973(7)     & \ $a_{00}$ \quad & 2.051(2) & 3.098(3) & 5.226(4) \\
$s_{01}$ & -82.27(1) & -69.01(1) & -49.356(8)   & $a_{01}$ & \ -15.033(3) & -12.851(3) & -9.426(3) \\
$s_{02}$ & 29.730(2) & 26.178(1) & 20.312(3)    & $a_{02}$ & -5.013(2) & -3.361(1) & -1.3526(7) \\
$s_{03}$ & -3.4399(7) & -2.9533(5) & -2.2166(3) & $a_{03}$ & 2.1103(3) & 1.7260(1) & 1.1251(2) \\
$s_{04}$ & -2.2750(4) & -1.6403(3) & -0.8547(2) & $a_{04}$ & 0.0434(2) & 0.01636(1) & \ -0.01074(5) \ \\
$s_{10}$ & -1854.4(2) & -1107.0(1) & -444.69(4) \ & $a_{10}$ & 43.75(1) & 36.66(1) & 25.827(9) \\
$s_{11}$ & 506.26(5) & 364.01(3) & 192.35(1)    & $a_{11}$ & 76.993(3) & 57.190(3) & 31.768(2) \\
$s_{12}$ & -95.42(2) & -70.94(1) & -40.162(6)   & $a_{12}$ & 44.260(4) & 29.363(2) & 12.962(1) \\
$s_{13}$ & 7.494(1) & 5.356(1) & 2.8030(4)      & $a_{13}$ & \ -4.4660(6) & -3.3740(5) & -1.8710(2) \\
$s_{20}$ & 18317(2) & 10081(1) & 3511.3(4)      & $a_{20}$ & -126.45(1) & -92.853(7) & -50.378(1) \\
$s_{21}$ & -2061.8(2) & -1350.7(1) & -595.79(7) \ & $a_{21}$ & -259.59(3) & -175.65(2) & -81.45(1) \\
$s_{22}$ & 202.75(7) & 133.19(4) & 59.25(2)     & $a_{22}$ & -107.48(1) & -67.737(8) & -27.500(3) \\
$s_{30}$ & -96390(10) & -50300(5) & -16185(2)   & $a_{30}$ & 295.76(3) & 198.78(2) & 90.96(1) \\
$s_{31}$ & 3784.8(4) & 2336.0(3) & 925.6(1)     & $a_{31}$ & 400.05(5) & 253.87(3) & 104.74(1) \\
$s_{40}$ & \ 213470(20) & 106940(10) & 32572(3) \  & $a_{40}$ & \ -348.41(4) & -220.12(3) & -90.11(1) \ \\
\hline
\end{tabular}
\caption{Numerical coefficients for the Scalar and Axial Vector operators using Wilson/clover fermions.} 
\label{tab2}
\end{center}
\end{table} 

For the Axial Vector operator we find:
\bea 
\ZA^{\rm singlet} (a \bar{\mu}) &-& \ZA^{\rm nonsinglet} (a \bar{\mu}) =  \nonumber \\
&-& \frac{g_\circ^4}{(4 \pi)^4} \ c_F \ N_f \ \Bigg\lbrace \ \MyOrange{\mathbf{6} \ \boldsymbol{\ln}(\boldsymbol{a^2 \bar{\mu}^2})} + (a_{00} + a_{01} \ \Red{\mathbf{c_{SW}}} + a_{02} \ \Red{\mathbf{c_{SW}^2}} + a_{03} \ \Red{\mathbf{c_{SW}^3}} + a_{04} \ \Red{\mathbf{c_{SW}^4}}) \nonumber \\
&+& (a_{10} + a_{11} \ \Red{\mathbf{c_{SW}}} + a_{12} \ \Red{\mathbf{c_{SW}^2}} + a_{13} \ \Red{\mathbf{c_{SW}^3}}) \ \Blue{\boldsymbol{\omega}} + (a_{20} + a_{21} \ \Red{\mathbf{c_{SW}}} + a_{22} \ \Red{\mathbf{c_{SW}^2}}) \ \Blue{\boldsymbol{\omega^2}} \nonumber \\
&+& (a_{30} + a_{31} \ \Red{\mathbf{c_{SW}}}) \ \Blue{\boldsymbol{\omega^3}} + a_{40} \ \Blue{\boldsymbol{\omega^4}} \Bigg\rbrace + \Op (g_\circ^6)
\label{ZA}
\eea
By analogy with the scalar case, the computation was performed in a general gauge and the
numerical constants $a_{ij}$ are tabulated in Table \ref{tab2}. We note that the result for the Axial Vector operator has a scale dependence; this is related to the axial anomaly.

\bigskip

Finally, the presence of a term of the form $\gamma_5 \ q_\mu \ \qslash / q^2$ in the Green's function of the
Axial Vector operator (Eq. \ref{GreensFunctionA}) implies that, in the alternative $RI'$ scheme mentioned
in Section II C, the above result is modified by a finite term, as below:
\be
\ZA^{\rm singlet (alter)} (a \bar{\mu}) - \ZA^{\rm nonsinglet (alter)} (a \bar{\mu}) = \ZA^{\rm singlet} (a \bar{\mu}) - \ZA^{\rm nonsinglet} (a \bar{\mu}) + \frac{g_\circ^4}{(4 \pi)^4} \ c_F \ N_f \label{ZAvdiff}
\ee

\subsection{Results from the staggered formulation of the fermion action}

\begin{figure}[htbp]
\centering
\epsfig{file=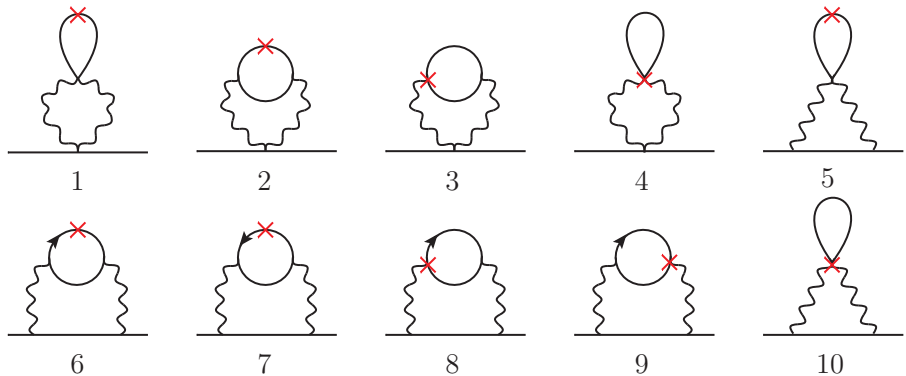,scale=0.75}
\caption{Diagrams (in staggered formulation) contributing to the difference between flavor singlet and nonsinglet values of $Z_\Gamma$. Solid (wavy) lines represent fermions (gluons). A cross denotes insertion of the operator ${\cal{O}}_\Gamma$.}
\label{figDiagS}
\end{figure}

In this subsection, we present the computational procedure and results 
using the staggered formulation of the fermion action (Eq. \ref{SFactionimpr1}). In this case, there are some additional vertices
with gluon lines stemming from the definition of bilinear operators (from the product $U_{C,D}$ in Eq. \ref{O_general}). As a result, there are 5 more Feynman diagrams, besides the 5 diagrams in Fig. \ref{figDiagW}, that enter our two-loop calculation. The extra diagrams involve operator vertices (the cross in the diagram) with up to two gluons. Fig. ~\ref{figDiagS}, shows a total of 10 two-loop Feynman diagrams contributing to the difference between singlet and nonsinglet Green's functions (Eqs. \ref{GreensFunctionS} - \ref{GreensFunctionT}). For $\Op_{\rm S}$ only diagrams 1, 2, 5, 6 and 7 contribute, since $U_{C,D} = \openone$. Similarly to the case of Wilson fermions, we have worked with $\sum_y \mathcal{O}_\Gamma (y)$. 

\bigskip

The above diagrams are more complicated than the corresponding diagrams in the Wilson case. One reason for this is the appearance of divergences in nontrivial corners of the Brillouin zone. Also, the presence of operator vertices with gluon lines, besides increasing the number of diagrams, gives terms with unusual offsets in momentum conservation delta functions (e.g., $\delta^{(4)}_{2 \pi} (p_1 + p_2 + \pi \hat{\mu})$); it turns out that these terms vanish in the final expression of each diagram. In addition, the two (rather than one) smearing steps of gluon links in the fermion action, as well as in the operators, lead to extremely lengthy vertices; the lengthiest cases which appeared in our computation are the operator vertices with two gluons ($\sim$ 800000 terms for the Axial Vector). Since these vertices appear only with their fermion lines contracted among themselves (diagrams: 4 and 10), we do not need to compute them as individual objects; we have used this fact in order to simplify the expression for diagrams 4 and 10. 

\bigskip

Another important issue for the above diagrams is exploiting the underlying symmetries. For example in the Wilson case, the denominator of the fermion propagator satisfies the symmetry $p_\mu \rightarrow -p_\mu , \forall \mu$; in the staggered case there is another symmetry: $p_\mu \rightarrow p_\mu + \pi \hat{\nu}$, where $\mu,\nu$ can be in the same or in different directions. This is a consequence of the semi-periodicity of the function $\sin^2 (p_\mu)$, which appears in the denominator of the staggered propagator (rather than $\sin^2 (p_\mu / 2)$). These symmetries help us to reduce the number of terms in the diagrams, eliminating odd integrands. 

\bigskip 

As in the Wilson case, the contribution of the diagrams in Fig. \ref{figDiagS} to $\ZP$, $\ZV$, $\ZT$ vanishes. Furthermore, in contrast to the Wilson case, $\ZS$ also receives a vanishing contribution. The closed fermion loop of the diagrams which contribute to $\ZS$, $\ZP$, $\ZT$, gives an odd number of exponentials of the inner momentum; this leads to odd integrands, which equal zero, due to the symmetry of the staggered propagator  mentioned above. So, for the cases of $\ZS$, $\ZP$, $\ZT$, the contribution vanishes diagram by diagram. Conversely, for the case of $\ZV$, each diagram vanishes when we add its symmetric diagram (diagrams 6+7, 8+9). Therefore, only $\ZA$ is affected. In particular, only diagrams 6 - 9 contribute to $\ZA$; the remaining diagrams vanish. Then, for the Axial Vector operator, our result can be written in the following form:
\bea 
\ZA^{\rm singlet} (a \bar{\mu}) &-& \ZA^{\rm nonsinglet} (a \bar{\mu}) =  \nonumber \\
&-& \frac{g_\circ^4}{(4 \pi)^4} \ c_F \ N_f \ \Bigg\lbrace \ \MyOrange{\mathbf{6} \ \boldsymbol{\ln}(\boldsymbol{a^2 \bar{\mu}^2})} + {\alpha}_1 + {\alpha}_2 \ (\Red{\boldsymbol{{\omega}_{A_1}}} + \Magenta{\boldsymbol{{\omega}_{A_2}}}) + {\alpha}_3 \ (\Red{\boldsymbol{{\omega}_{A_1}^2}} + \Magenta{\boldsymbol{{\omega}_{A_2}^2}}) + {\alpha}_4 \ \Red{\boldsymbol{{\omega}_{A_1}}} \ \Magenta{\boldsymbol{{\omega}_{A_2}}} \nonumber \\
&+& {\alpha}_5 \ (\Red{\boldsymbol{{\omega}_{A_1}^3}} + \Magenta{\boldsymbol{{\omega}_{A_2}^3}}) + {\alpha}_6 \ \Red{\boldsymbol{{\omega}_{A_1}}} \ \Magenta{\boldsymbol{{\omega}_{A_2}}} \ (\Red{\boldsymbol{{\omega}_{A_1}}} + \Magenta{\boldsymbol{{\omega}_{A_2}}}) + {\alpha}_7 \ (\Red{\boldsymbol{{\omega}_{A_1}^4}} + \Magenta{\boldsymbol{{\omega}_{A_2}^4}}) + {\alpha}_8 \ \Red{\boldsymbol{{\omega}_{A_1}^2}} \ \Magenta{\boldsymbol{{\omega}_{A_2}^2}} \nonumber \\
&+& {\alpha}_9 \ \Red{\boldsymbol{{\omega}_{A_1}}} \ \Magenta{\boldsymbol{{\omega}_{A_2}}} \ (\Red{\boldsymbol{{\omega}_{A_1}^2}} + \Magenta{\boldsymbol{{\omega}_{A_2}^2}}) + {\alpha}_{10} \ \Red{\boldsymbol{{\omega}_{A_1}^2}} \ \Magenta{\boldsymbol{{\omega}_{A_2}^2}} \ (\Red{\boldsymbol{{\omega}_{A_1}}} + \Magenta{\boldsymbol{{\omega}_{A_2}}}) + {\alpha}_{11} \ \Red{\boldsymbol{{\omega}_{A_1}}} \ \Magenta{\boldsymbol{{\omega}_{A_2}}} \ (\Red{\boldsymbol{{\omega}_{A_1}^3}} + \Magenta{\boldsymbol{{\omega}_{A_2}^3}}) \nonumber \\
&+& {\alpha}_{12} \ \Red{\boldsymbol{{\omega}_{A_1}^3}} \ \Magenta{\boldsymbol{{\omega}_{A_2}^3}} + {\alpha}_{13} \ \Red{\boldsymbol{{\omega}_{A_1}^2}} \ \Magenta{\boldsymbol{{\omega}_{A_2}^2}} \ (\Red{\boldsymbol{{\omega}_{A_1}^2}} + \Magenta{\boldsymbol{{\omega}_{A_2}^2}}) + {\alpha}_{14} \ \Red{\boldsymbol{{\omega}_{A_1}^3}} \ \Magenta{\boldsymbol{{\omega}_{A_2}^3}} \ (\Red{\boldsymbol{{\omega}_{A_1}}} + \Magenta{\boldsymbol{{\omega}_{A_2}}}) \nonumber \\ 
&+& {\alpha}_{15} \ \Red{\boldsymbol{{\omega}_{A_1}^4}} \ \Magenta{\boldsymbol{{\omega}_{A_2}^4}} + {\alpha}_{16} \ (\Blue{\boldsymbol{{\omega}_{O_1}}} + \Cyan{\boldsymbol{{\omega}_{O_2}}}) + {\alpha}_{17} \ \Blue{\boldsymbol{{\omega}_{O_1}}} \ \Cyan{\boldsymbol{{\omega}_{O_2}}} + {\alpha}_{18} \ (\Red{\boldsymbol{{\omega}_{A_1}}} + \Magenta{\boldsymbol{{\omega}_{A_2}}}) \  (\Blue{\boldsymbol{{\omega}_{O_1}}} + \Cyan{\boldsymbol{{\omega}_{O_2}}}) \nonumber \\
&+& {\alpha}_{19} \ \Red{\boldsymbol{{\omega}_{A_1}}} \ \Magenta{\boldsymbol{{\omega}_{A_2}}} \ (\Blue{\boldsymbol{{\omega}_{O_1}}} + \Cyan{\boldsymbol{{\omega}_{O_2}}}) + {\alpha}_{20} \ \Big[(\Red{\boldsymbol{\omega_{A_1}^2}} + \Magenta{\boldsymbol{\omega_{A_2}^2}}) \ (\Blue{\boldsymbol{\omega_{O_1}}} + \Cyan{\boldsymbol{\omega_{O_2}}}) + (\Red{\boldsymbol{\omega_{A_1}}} + \Magenta{\boldsymbol{\omega_{A_2}}}) \ \Blue{\boldsymbol{\omega_{O_1}}} \ \Cyan{\boldsymbol{\omega_{O_2}}} \Big] \nonumber \\
&+& {\alpha}_{21} \ (\Red{\boldsymbol{\omega_{A_1}^2}} + \Magenta{\boldsymbol{\omega_{A_2}^2}}) \ \Blue{\boldsymbol{\omega_{O_1}}} \ \Cyan{\boldsymbol{\omega_{O_2}}} + {\alpha}_{22} \ (\Red{\boldsymbol{\omega_{A_1}^3}} + \Magenta{\boldsymbol{\omega_{A_2}^3}}) \ (\Blue{\boldsymbol{\omega_{O_1}}} + \Cyan{\boldsymbol{\omega_{O_2}}}) \nonumber \\
&+& {\alpha}_{23} \ \Red{\boldsymbol{\omega_{A_1}}} \ \Magenta{\boldsymbol{\omega_{A_2}}} \ \Big[ (\Red{\boldsymbol{\omega_{A_1}}} + \Magenta{\boldsymbol{\omega_{A_2}}}) \ (\Blue{\boldsymbol{\omega_{O_1}}} + \Cyan{\boldsymbol{\omega_{O_2}}}) + \Blue{\boldsymbol{\omega_{O_1}}} \ \Cyan{\boldsymbol{\omega_{O_2}}}\Big] + {\alpha}_{24} \ (\Red{\boldsymbol{\omega_{A_1}^3}} + \Magenta{\boldsymbol{\omega_{A_2}^3}}) \  \Blue{\boldsymbol{\omega_{O_1}}} \ \Cyan{\boldsymbol{\omega_{O_2}}} \nonumber \\
&+& {\alpha}_{25} \ \Red{\boldsymbol{\omega_{A_1}}} \ \Magenta{\boldsymbol{\omega_{A_2}}} \ (\Red{\boldsymbol{\omega_{A_1}^2}} + \Magenta{\boldsymbol{\omega_{A_2}^2}}) \ (\Blue{\boldsymbol{\omega_{O_1}}} + \Cyan{\boldsymbol{\omega_{O_2}}}) \nonumber \\
&+& {\alpha}_{26} \ \Red{\boldsymbol{\omega_{A_1}}} \ \Magenta{\boldsymbol{\omega_{A_2}}} \ \Big[\Red{\boldsymbol{\omega_{A_1}}} \ \Magenta{\boldsymbol{\omega_{A_2}}} \ (\Blue{\boldsymbol{\omega_{O_1}}} + \Cyan{\boldsymbol{\omega_{O_2}}}) + (\Red{\boldsymbol{\omega_{A_1}}} + \Magenta{\boldsymbol{\omega_{A_2}}}) 
 \Blue{\boldsymbol{\omega_{O_1}}} \ \Cyan{\boldsymbol{\omega_{O_2}}} \Big] + {\alpha}_{27} \ \Red{\boldsymbol{\omega_{A_1}^2}} \ \Magenta{\boldsymbol{\omega_{A_2}^2}} \ \Blue{\boldsymbol{\omega_{O_1}}} \ \Cyan{\boldsymbol{\omega_{O_2}}} \nonumber \\
&+& {\alpha}_{28} \ \Red{\boldsymbol{\omega_{A_1}}} \ \Magenta{\boldsymbol{\omega_{A_2}}} \ \Big[\Red{\boldsymbol{\omega_{A_1}}} \ \Magenta{\boldsymbol{\omega_{A_2}}} \ (\Red{\boldsymbol{\omega_{A_1}}} + \Magenta{\boldsymbol{\omega_{A_2}}}) \ (\Blue{\boldsymbol{\omega_{O_1}}} + \Cyan{\boldsymbol{\omega_{O_2}}}) + (\Red{\boldsymbol{\omega_{A_1}^2}} + \Magenta{\boldsymbol{\omega_{A_2}^2}}) \ \Blue{\boldsymbol{\omega_{O_1}}} \ \Cyan{\boldsymbol{\omega_{O_2}}} \Big] \nonumber \\
&+& {\alpha}_{29} \ \Red{\boldsymbol{\omega_{A_1}^3}} \ \Magenta{\boldsymbol{\omega_{A_2}^3}} \ (\Blue{\boldsymbol{\omega_{O_1}}} + \Cyan{\boldsymbol{\omega_{O_2}}}) + {\alpha}_{30} \ \Red{\boldsymbol{\omega_{A_1}^2}} \ \Magenta{\boldsymbol{\omega_{A_2}^2}} \ (\Red{\boldsymbol{\omega_{A_1}}} + \Magenta{\boldsymbol{\omega_{A_2}}}) \ \Blue{\boldsymbol{\omega_{O_1}}} \ \Cyan{\boldsymbol{\omega_{O_2}}} \nonumber \\
&+& {\alpha}_{31} \ \Red{\boldsymbol{\omega_{A_1}^3}} \ \Magenta{\boldsymbol{\omega_{A_2}^3}} \ \Blue{\boldsymbol{\omega_{O_1}}} \ \Cyan{\boldsymbol{\omega_{O_2}}} \Bigg\rbrace + \Op (g_\circ^6)
\label{ZA_stag}
\eea
The computation was performed in a general gauge and the numerical constants $\alpha_i$ are tabulated in Table \ref{tab3}. 
\begin{table}[!ht]
\begin{center}
\begin{tabular}{|c|ccc||c|ccc|}
\hline
 & Wilson & TL Symanzik & Iwasaki & & Wilson & TL Symanzik & Iwasaki  \\
\hline
\ ${\alpha}_{1}$ \quad & 17.420(1) & 16.000(1) & 14.610(1)     & \ ${\alpha}_{16}$ \quad & 24.9873(2) & 18.0489(4) & 9.9571(2) \\
\ ${\alpha}_{2}$ & -116.049(7) & -81.342(5) & -41.583(2)   & \ ${\alpha}_{17}$ & -97.4550(2) & -62.2675(1) & -26.5359(1) \\
\ ${\alpha}_{3}$ & 839.788(9) & 539.121(6) & 230.050(1)    & \ ${\alpha}_{18}$ & -292.3650(5) & -186.8025(4) & -79.6078(2) \\
\ ${\alpha}_{4}$ & 2175.14(3) & 1394.12(2) & 591.88(1) & \ ${\alpha}_{19}$ & 4864.513(9) & 2921.876(6) & 1107.333(2) \\
\ ${\alpha}_{5}$ & -3462.830(1) & -2098.136(5) & -801.633(3) & \ ${\alpha}_{20}$ & 1621.504(3) & 973.959(2) & 369.111(1) \\
\ ${\alpha}_{6}$ & -19565.9(1) & -11858.6(1) & -4528.6(1) & \ ${\alpha}_{21}$ & -10617.81(2) & -6122.11(1) & -2169.30(1) \\
\ ${\alpha}_{7}$ & 6424.33(2) & 3740.18(1) & 1337.93(1)    & \ ${\alpha}_{22}$ & -3539.269(6) & -2040.705(4) & -723.099(1) \\
\ ${\alpha}_{8}$ & 200966.5(4) & 117179.7(4) & 41977.1(1)   & \ ${\alpha}_{23}$ & -31853.42(5) & -18366.34(3) & -6507.89(1) \\
\ ${\alpha}_{9}$ & 92171.5(3) & 53720.8(1) & 19237.6(1)      & \ ${\alpha}_{24}$ & 25847.14(3) & 14435.59(2) & 4881.52(1) \\
\ ${\alpha}_{10}$ & -1026448(1) & -580271(2) & -198722(1)      & \ ${\alpha}_{25}$ & 77541.41(1) & 43306.78(6)  & 14644.54(2) \\
\ ${\alpha}_{11}$ & -183998.3(3) & -103929.7(3) & -35561.1(1) & \ ${\alpha}_{26}$ & 232624.2(3) & 129920.3(2) & 43933.6(1) \\
\ ${\alpha}_{12}$ & 5517230(30) &  3037110(10) & 1003641(1)     & \ ${\alpha}_{27}$ & -1844375(1) & -1002465(1) & -326727(1) \\
\ ${\alpha}_{13}$ & 2145810(10) & 1180684(4) & 389979(1)   & \ ${\alpha}_{28}$ & \ -614791.6(6) & -334155.0(4) & -108909.0(2) \ \\
\ ${\alpha}_{14}$ & \ -11889300(40) & -6386950(30) & -2046240(10) \    & \ ${\alpha}_{29}$ & \ 1736048.1(8) & 920956.7(7) & 290916.1(3) \ \\
\ ${\alpha}_{15}$ & \ 26137700(200) & 13729010(10) & 4278680(10) \  & \ ${\alpha}_{30}$ & 5208144(2) & 2762870(2) & 872748(1) \\
                &            &            &            & \ ${\alpha}_{31}$ & \ -15545543(1) & -8065557(2) & -2478207(1) \ \\
\hline
\end{tabular}
\caption{Numerical coefficients for the Axial Vector operator using staggered fermions.} 
\label{tab3}
\end{center}
\end{table}
The computation with staggered fermions gives rise to some nontrivial divergent integrals, which cannot be present in the Wilson formulation due to the different pole structure of the fermion propagator. In Appendix B, we provide a brief description of the manipulations
performed to evaluate such divergent terms. We note that the result for the Axial Vector operator, as we expected, has the same divergent behaviour just as in the Wilson case and in the continuum, i.e. $6 \ln (a^2 \bar{\mu}^2)$. As was expected, this logarithmic divergence originates in diagrams 6 and 7, which are the only ones present in the continuum. Also, in the alternative $RI'$ scheme we must add the same finite term,  $g_\circ^4 / (4 \pi)^4 \ c_F \ N_f$, as in the Wilson case. 

\newpage

\section{Discussion}

The numerical value of the difference between singlet and nonsinglet renormalization functions can be very significant, depending on the values of the parameters employed in the action. In order to assess the importance of this difference, we present here several graphs of the results for certain values of $c_i, \ c_{SW}$ and $\omega$ for the Wilson case and $c_i, \ \omega_{A_1}, \ \omega_{A_2}, \ \omega_{O_1}$ and $\omega_{O_2}$ for the staggered case.

\bigskip

\begin{figure}[!ht]
\begin{center}
\epsfig{file=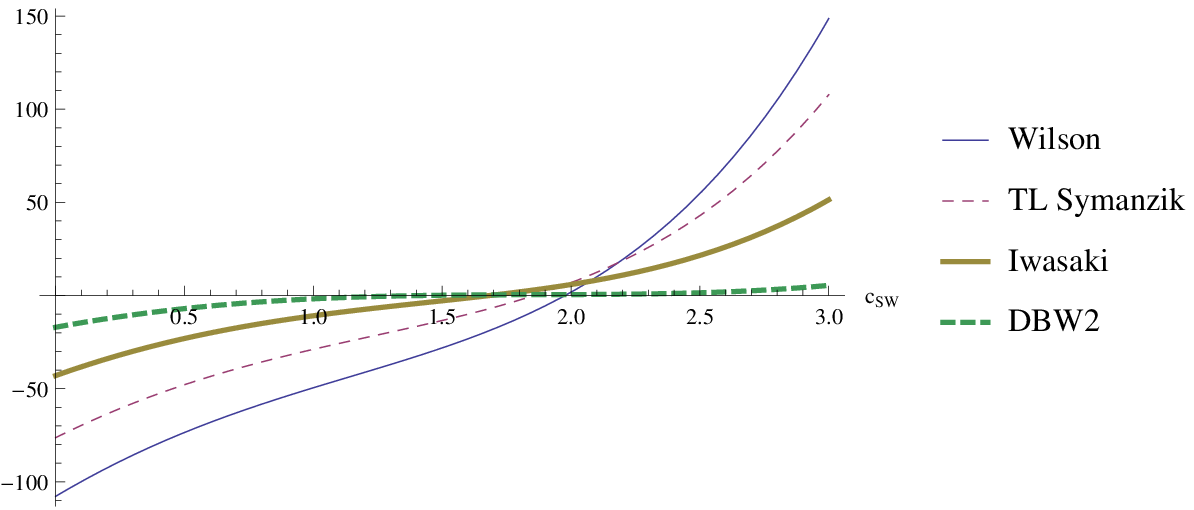,scale=0.6}
\hspace{.05\textwidth}
\epsfig{figure=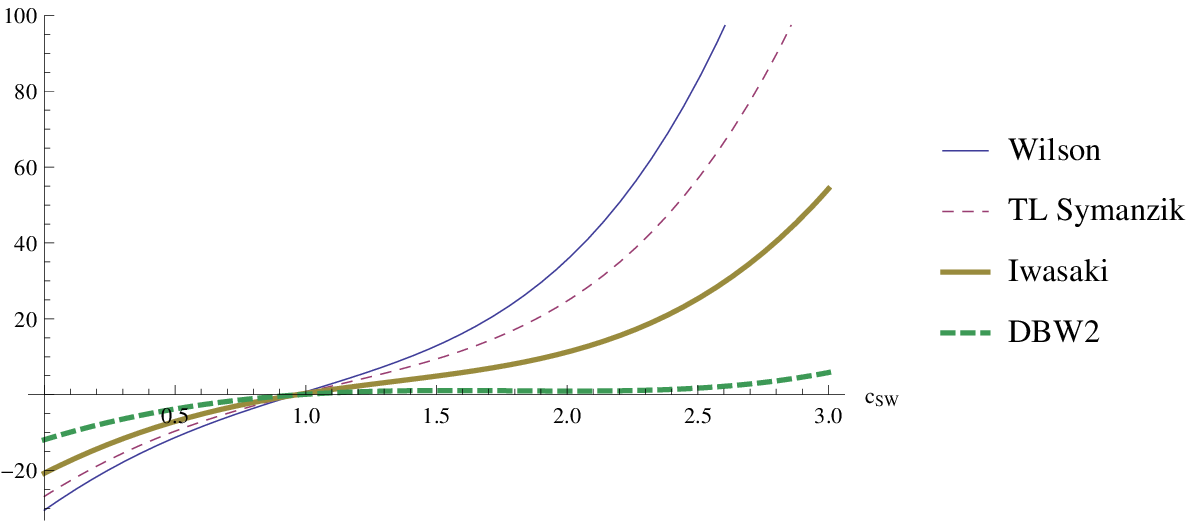,scale=0.6}\\[2mm]
\hspace{.004\textwidth}
\epsfig{figure=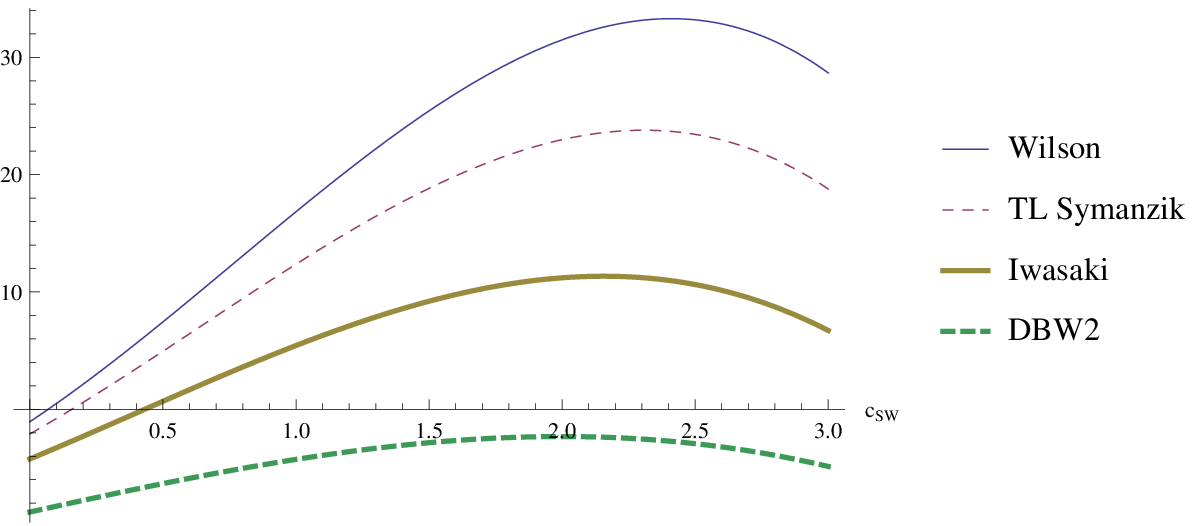,scale=0.6}
\hspace{.04\textwidth}
\epsfig{figure=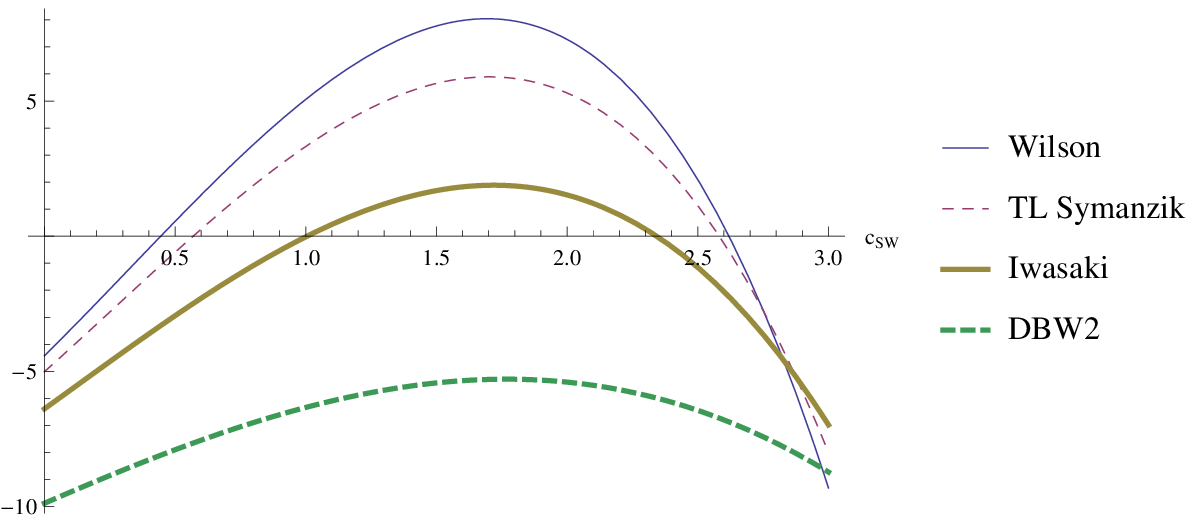,scale=0.6}
\caption{Plots of $\zeta \equiv \left[ Z_{\Gamma}^{\rm{(singlet)}} - Z_{\Gamma}^{\rm{(nonsinglet)}} \right] \left( \frac{g_o^4}{\left( 4 \pi \right)^4} N_f c_F \right)^{-1}$,
  for Scalar $\Gamma = S$ (top panels) and Axial Vector $\Gamma = A$ (bottom panels), as a function of
  $c_{SW}$, for $\omega = 0$ (left panels) and $\omega = 0.1$ (right panels).}
\label{ZVSCswplots}
\end{center}
\end{figure}

\begin{figure}[!ht]
\begin{center}
\epsfig{file=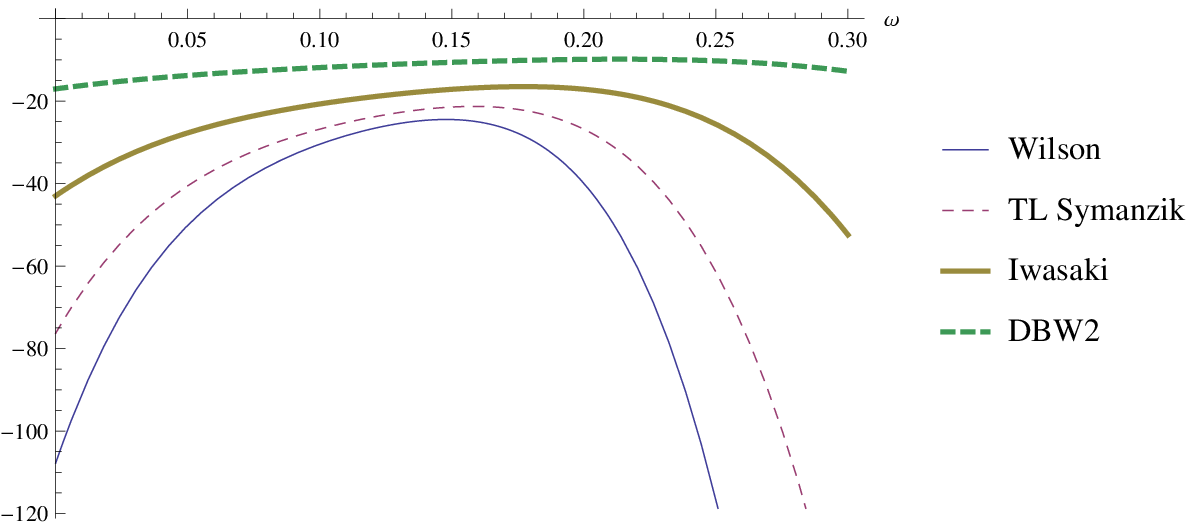,scale=0.6}
\hspace{.05\textwidth}
\epsfig{figure=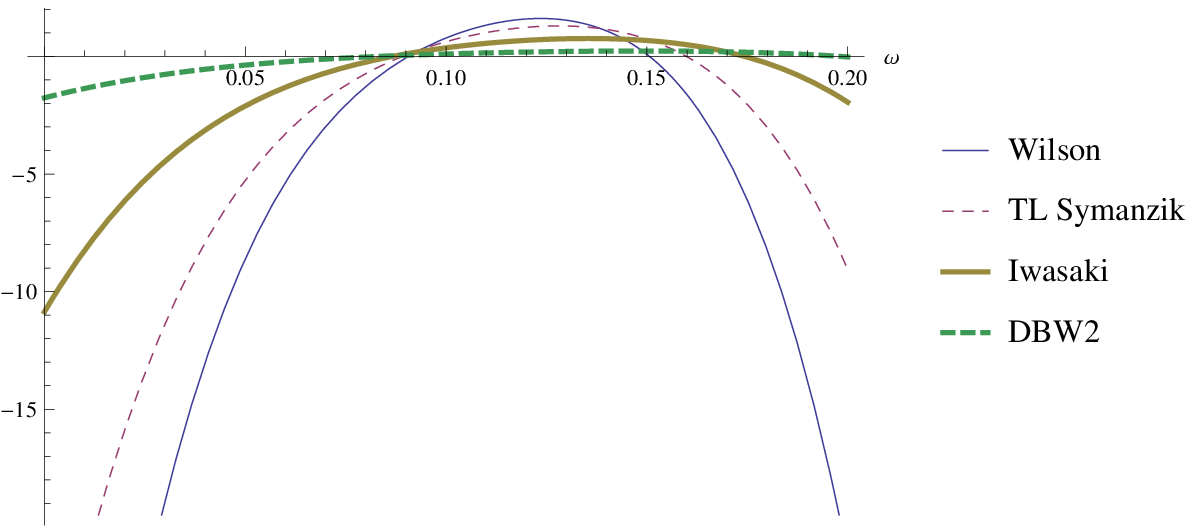,scale=0.6}\\[2mm]
\hspace{.001\textwidth}
\epsfig{figure=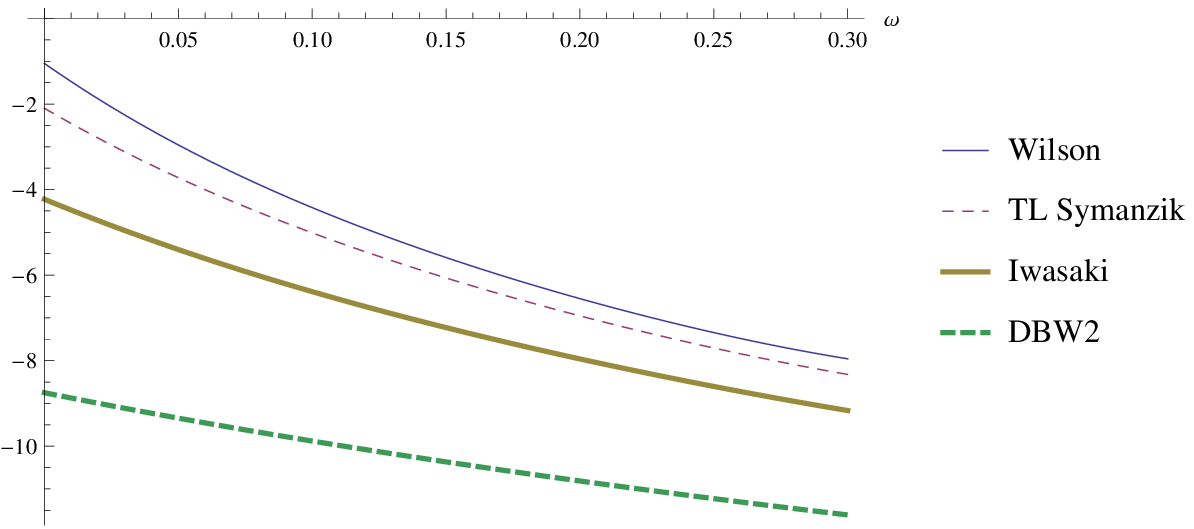,scale=0.6}
\hspace{.05\textwidth}
\epsfig{figure=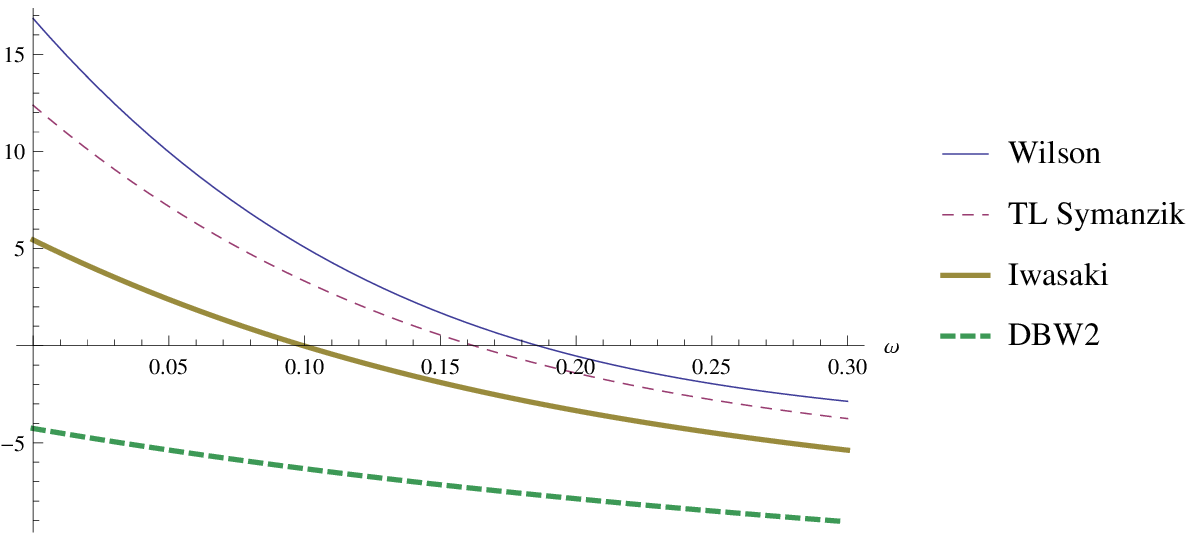,scale=0.6}
\caption{Plots of $\zeta \equiv \left[ Z_{\Gamma}^{\rm{(singlet)}} - Z_{\Gamma}^{\rm{(nonsinglet)}} \right] \left( \frac{g_o^4}{\left( 4 \pi \right)^4} N_f c_F \right)^{-1}$,
  for Scalar $\Gamma = S$ (top panels) and Axial Vector $\Gamma = A$ (bottom panels), as a function of
  $\omega$, for $c_{SW} = 0$ (left panels) and $c_{SW} = 1$ (right panels).}
\label{ZVSOmegaplots}
\end{center}
\end{figure}

In Figs. (\ref{ZVSCswplots} - \ref{ZVSOmegaplots}) we illustrate our results from the Wilson formulation by selecting values for parameters $c_{SW}$ and $\omega$. 
We compute the contribution to $\zeta$, defined through:
\be
 Z_{\Gamma}^{\rm{(singlet)}} - Z_{\Gamma}^{\rm{(nonsinglet)}} = \frac{g_o^4}{\left( 4 \pi \right)^4} N_f c_F \ \zeta
\ee 
using different gluon actions: Wilson, tree-level Symanzik, Iwasaki, DBW2. We notice that the Iwasaki and DBW2 actions exhibit a milder dependence altogether on $c_{SW}$ and $\omega$. Certain values of $c_{SW}$ and $\omega$ lead to almost vanishing contributions to $\zeta$ in the Scalar case (and less so in the Axial Vector case) for all gluon actions considered (see right panels of Fig. \ref{ZVSCswplots}). 

\bigskip

\begin{figure}[!ht]
\begin{center}
\vspace{0.05\textwidth}
\epsfig{file=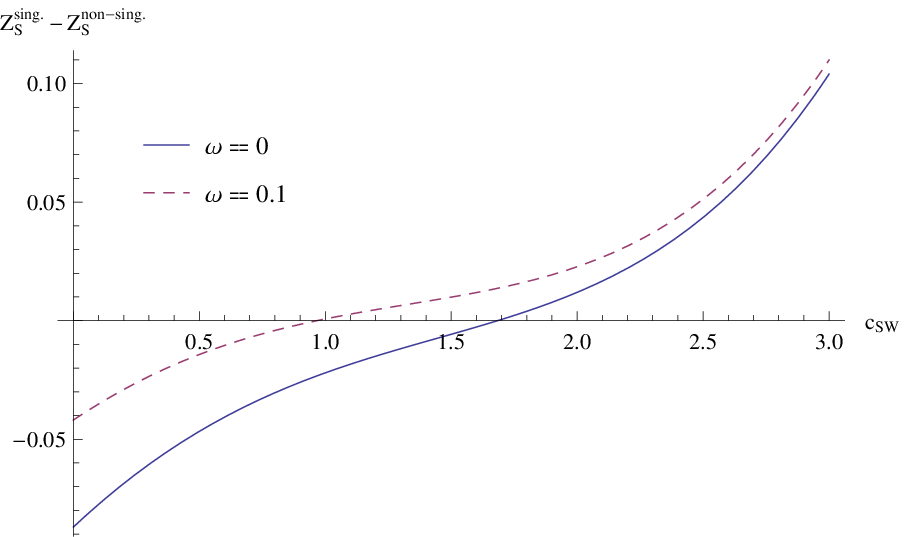,scale=0.7}
\hspace{.07\textwidth}
\epsfig{figure=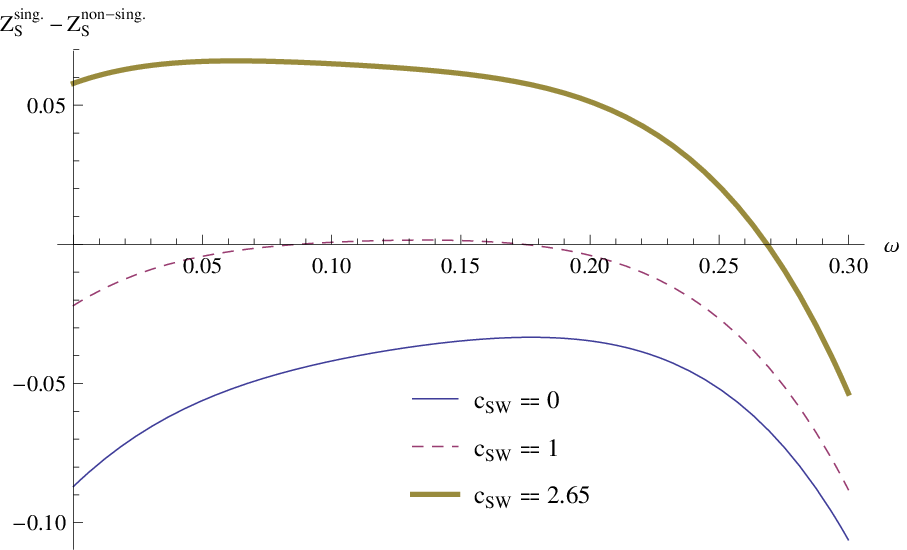,scale=0.7}\\[2mm]
\epsfig{figure=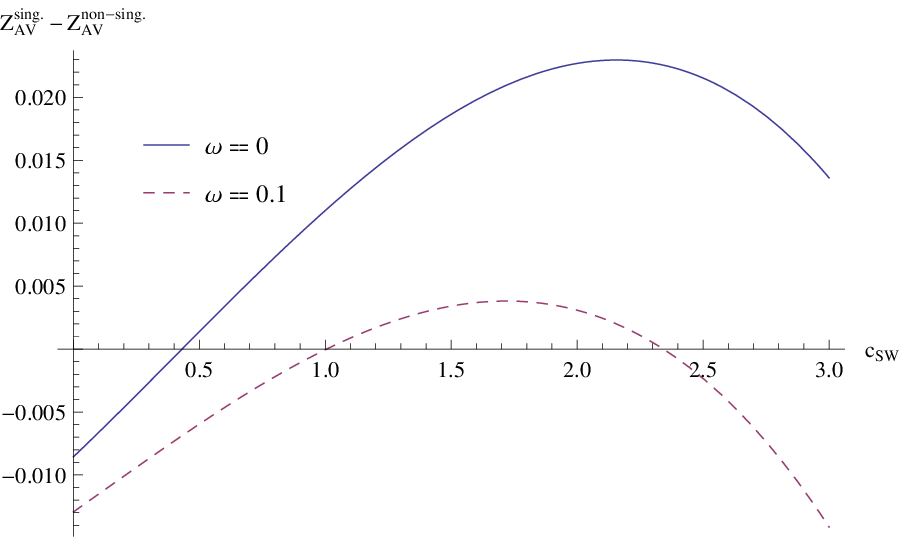,scale=0.7}
\hspace{.07\textwidth}
\epsfig{figure=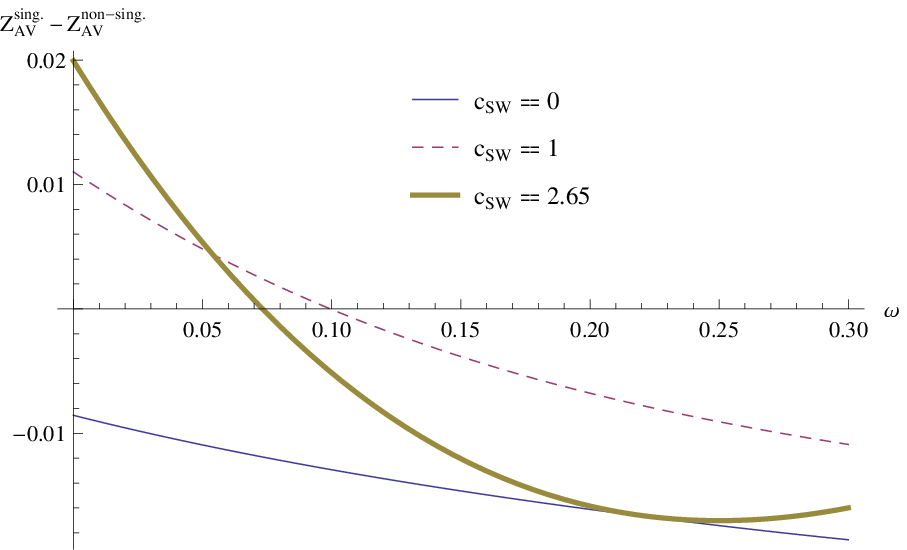,scale=0.7}
\caption{Plots of $Z_\Gamma^{\rm singlet} - Z_\Gamma^{\rm nonsiglet}$,
  for Scalar $\Gamma = S$ (top panels) and Axial Vector $\Gamma = AV$ (bottom), as a function of
  $c_{SW}$ (left) and $\omega$ (right). Parameter values relevant for
  ETMC action ($N_f = 4$, Iwasaki gluons, $\beta = 1.95$).}
\label{ETMCplots}
\end{center}
\end{figure}

\begin{figure}[!ht]
\begin{center}
\epsfig{figure=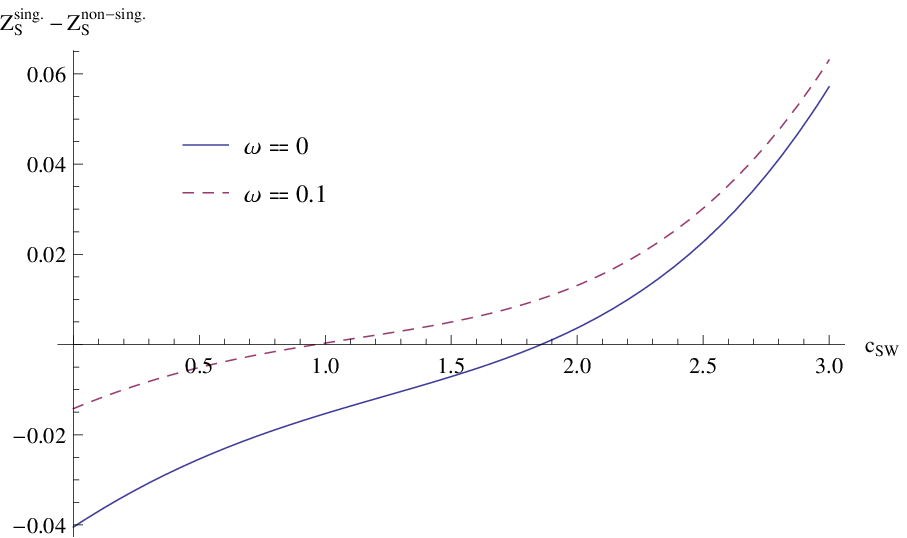,scale=0.7}
\hspace{.07\textwidth}
\epsfig{figure=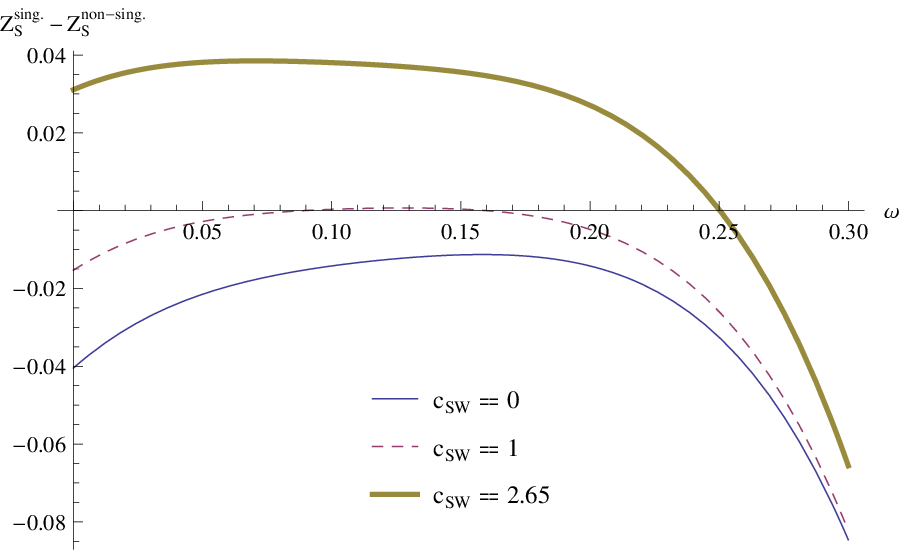,scale=0.7}\\[2mm]
\epsfig{figure=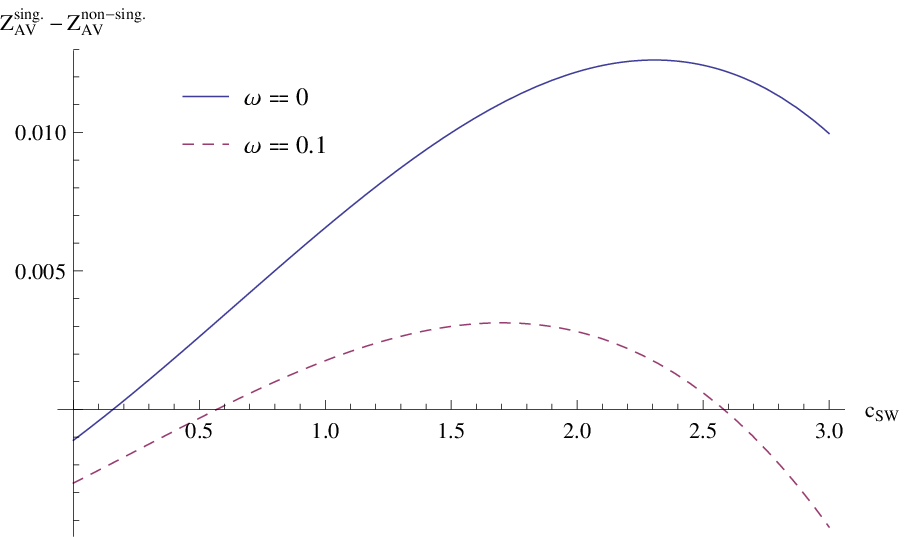,scale=0.7}
\hspace{.07\textwidth}
\epsfig{figure=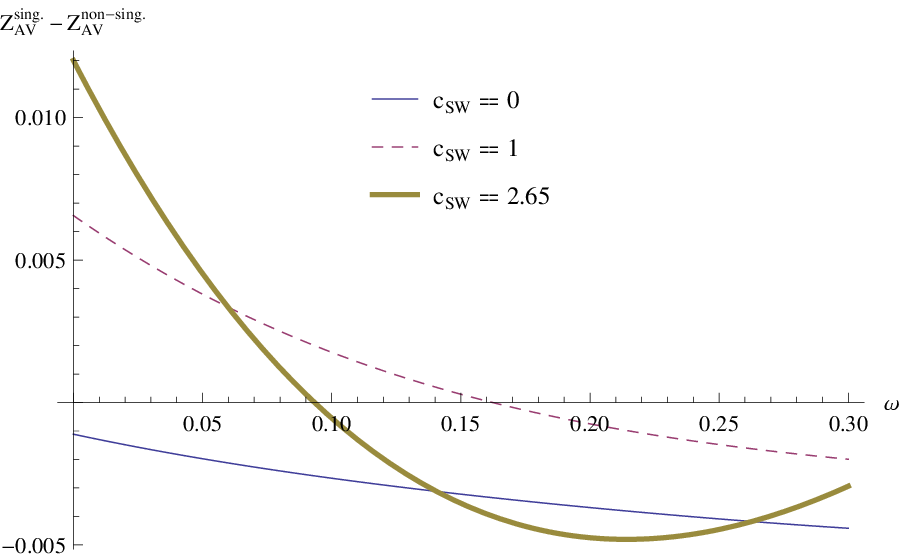,scale=0.7}
\caption{Plots of $Z_\Gamma^{\rm singlet} - Z_\Gamma^{\rm nonsiglet}$,
  for Scalar $\Gamma = S$ (top panels) and Axial Vector $\Gamma = AV$ (bottom), as a function of
  $c_{SW}$ (left) and $\omega$ (right). Parameter values relevant for
  SLiNC action ($N_f = 3$, TL Symanzik gluons, $\beta = 5.5$).}
\label{SLiNCplots}
\end{center}
\end{figure}

In Fig.~\ref{ETMCplots} we have selected
parameter values appropriate to the
ETMC action with Iwasaki gluons, $N_f = 2+1+1$, $\beta = 2 N_c/g_\circ^2 = 1.95$,
$\bar\mu = 1/a$ and standard/stout links for the Wilson part of the
fermion action. Results from this work have been successfully applied by the ETM Collaboration
for the renormalization of the individual quark contributions to the intrinsic spin~\cite{Constantinou:2014tga}.
We anticipate further use of our results for other action parameters in the near future.
Fig.~\ref{SLiNCplots} presents our results for
parameter values appropriate to the SLiNC action, with tree-level
Symanzik gluons, $N_f = 3$,  $\beta = 2 N_c \ c_0/g_\circ^2 =
5.5$, $\bar\mu = 1/a$\,.

\clearpage

In the staggered formulation, we note that the result \eqref{ZA_stag} is symmetric under the exchange of $\omega_{A}$s as well as under the exchange of $\omega_{O}$s. This fact is consistent with the requirement that the results for ($\omega_{O_1} = 0$, $\omega_{O_2} = \omega$) and ($\omega_{O_1} = \omega$, $\omega_{O_2} = 0$) should coincide, since they both correspond to a single smearing step; similarly for the coefficients $\omega_{A_1}$ and $\omega_{A_2}$. These properties provide nontrivial consistency checks of our computation. 

\bigskip

\begin{figure}[!ht]
\begin{center}
\epsfig{file=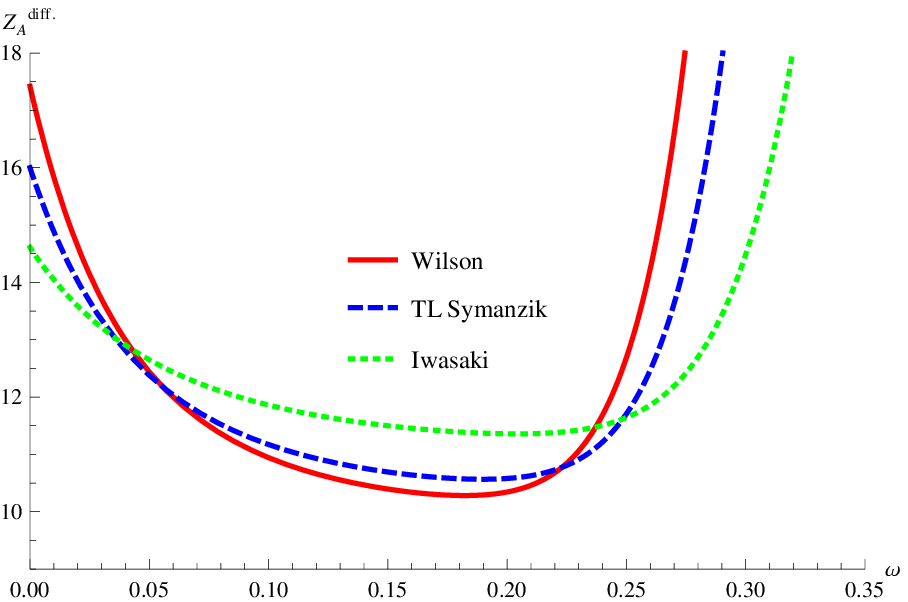,scale=0.75}
\hspace{.07\textwidth}
\epsfig{figure=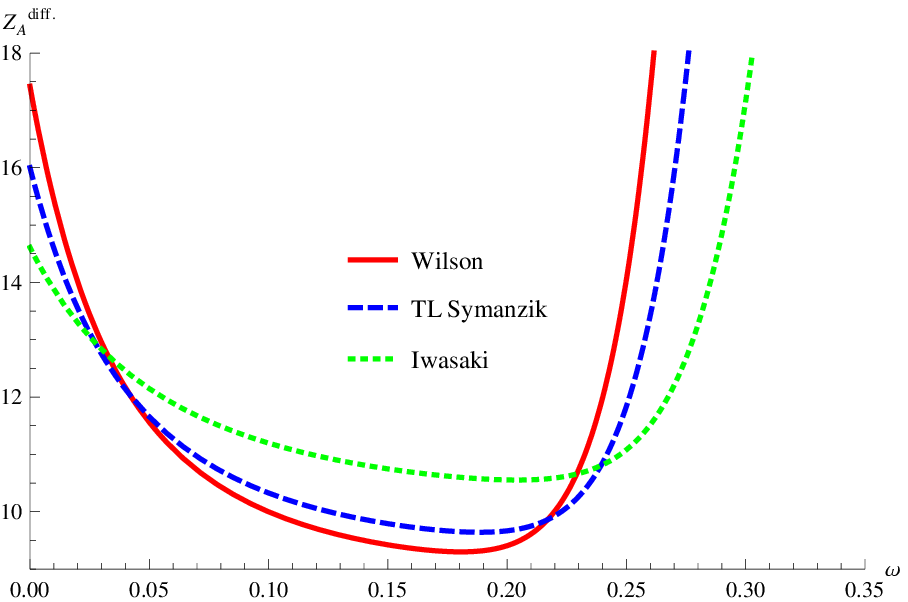,scale=0.75}\\[2mm]
\epsfig{figure=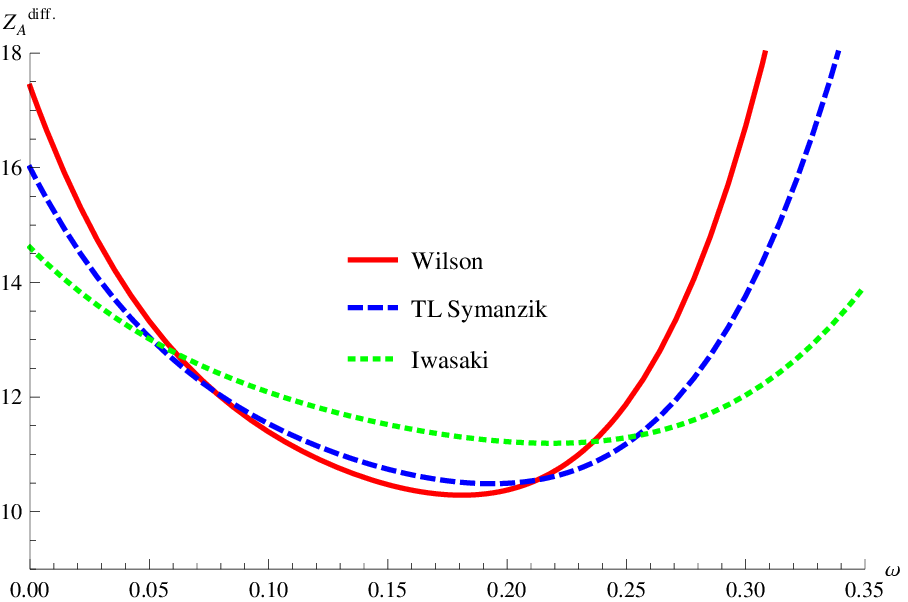,scale=0.75}
\hspace{.07\textwidth}
\epsfig{figure=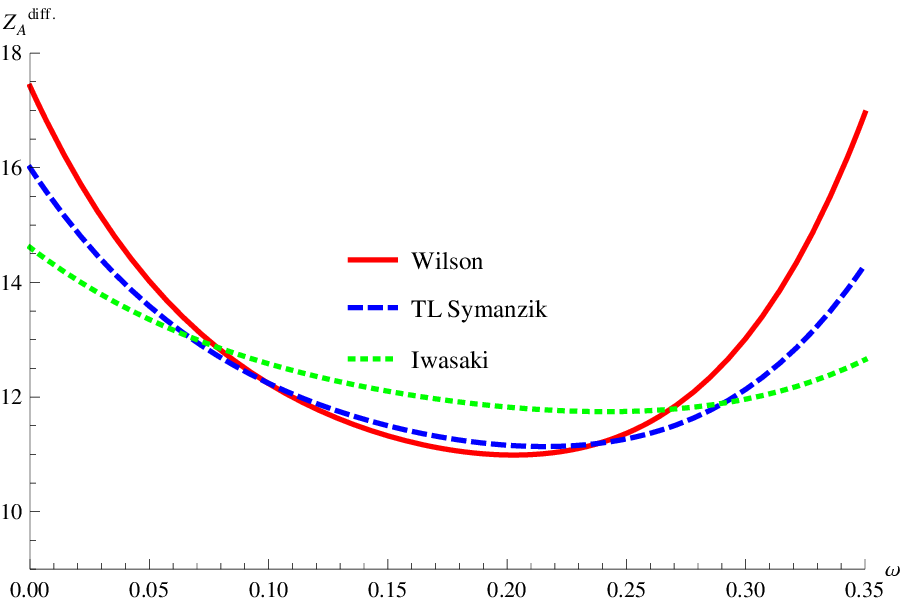,scale=0.75}
\caption{Plots of $Z_A^{\rm{diff.}} \equiv \left[ Z_A^{\rm{(singlet)}} - Z_A^{\rm{(nonsinglet)}} \right] \left( - \frac{g_o^4}{\left( 4 \pi \right)^4} N_f c_F \right)^{-1}$, as a function of $\omega$ for the parameter values: upper left: $\omega_{A_1} = \omega_{A_2} = \omega_{O_1} = \omega_{O_2} = \omega$, upper right: $\omega_{A_1} = \omega_{A_2} = \omega$, $\omega_{O_1} = \omega_{O_2} = 0$, lower left: $\omega_{A_1} = \omega$, $\omega_{A_2} = \omega_{O_1} = \omega_{O_2} = 0$, lower right: $\omega_{A_1} = \omega_{O_1} = \omega$, $\omega_{A_2} = \omega_{O_2} = 0$.}
\label{Stagplots2D}
\end{center}
\end{figure}

In Fig. \ref{Stagplots2D} we present 2D graphs of our results by selecting
the following parameter values: \\
\hspace*{0.75cm} 1. $\omega_{A_1} = \omega_{A_2} = \omega_{O_1} = \omega_{O_2} = \omega$ \\
\hspace*{0.75cm} 2. $\omega_{A_1} = \omega_{A_2} = \omega$, $\omega_{O_1} = \omega_{O_2} = 0$ (No smearing procedure in the links of operators) \\
\hspace*{0.75cm} 3. $\omega_{A_1} = \omega$, $\omega_{A_2} = \omega_{O_1} = \omega_{O_2} = 0$ (One smearing step only in the links of fermion action) \\
\hspace*{0.75cm} 4. $\omega_{A_1} = \omega_{O_1} = \omega$, $\omega_{A_2} = \omega_{O_2} = 0$ (One smearing step in the links of fermion action and operators). \\
The vertical axis of theses plots corresponds to
$Z_A^{\rm{diff.}} \equiv \left[ Z_A^{\rm{(singlet)}}\! \left( a \bar{\mu} \right) - Z_A^{\rm{(nonsinglet)}}\! \left( a \bar{\mu} \right) \right] \left( - \frac{g_o^4}{\left( 4 \pi \right)^4} N_f c_F \right)^{-1}$ for $\bar{\mu} = 1 / a$. We plot the results for gluon actions: Wilson, tree-level Symanzik, Iwasaki in the same graph. We notice that the plots for the Iwasaki action are flatter than for the remaining actions but the Wilson action has the smallest values of $Z_A^{\rm diff}$.

\bigskip

\begin{figure}[!ht]
\vspace*{-0.5cm}
\begin{center}
\epsfig{file=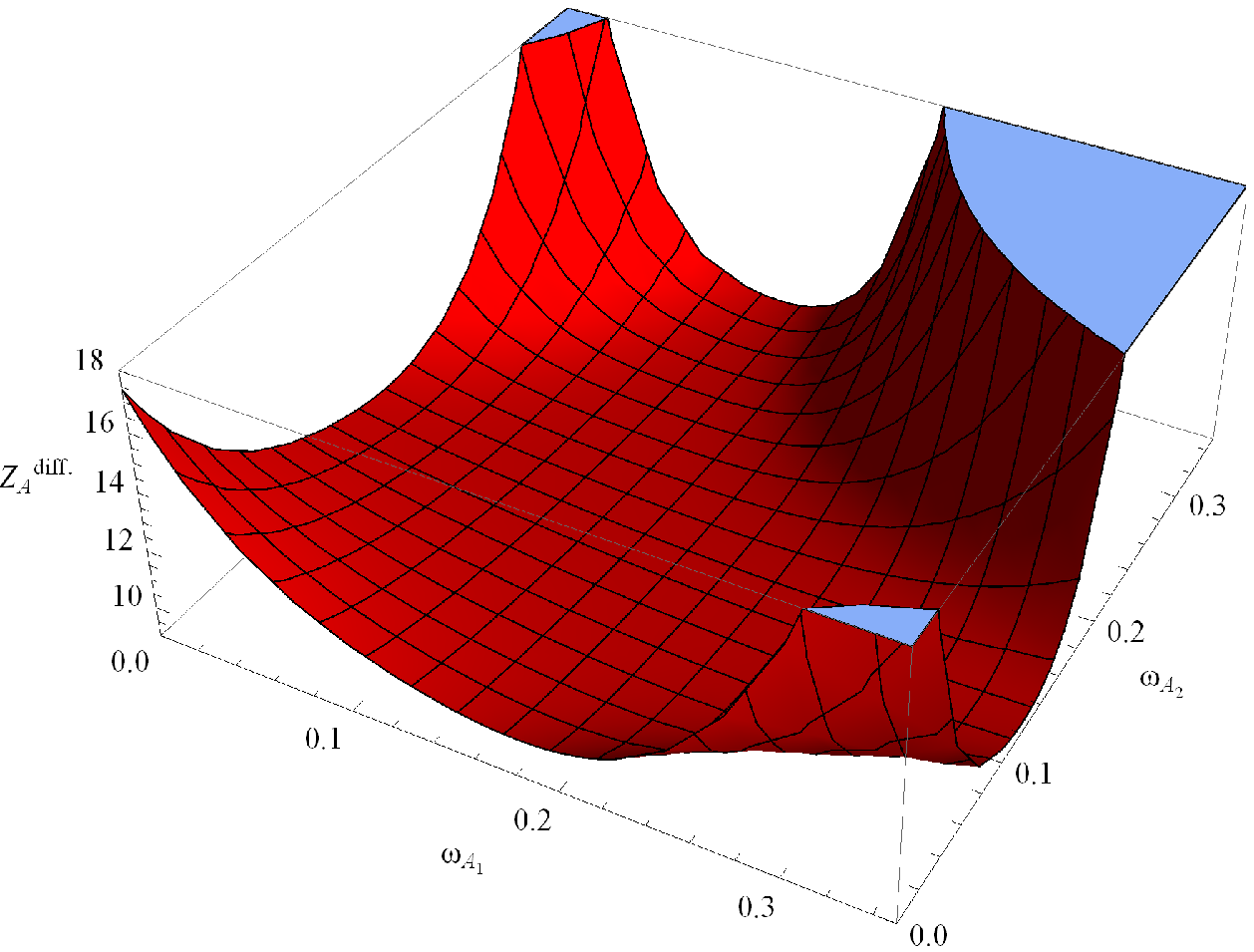,scale=0.5}
\hspace{.05\textwidth}
\epsfig{figure=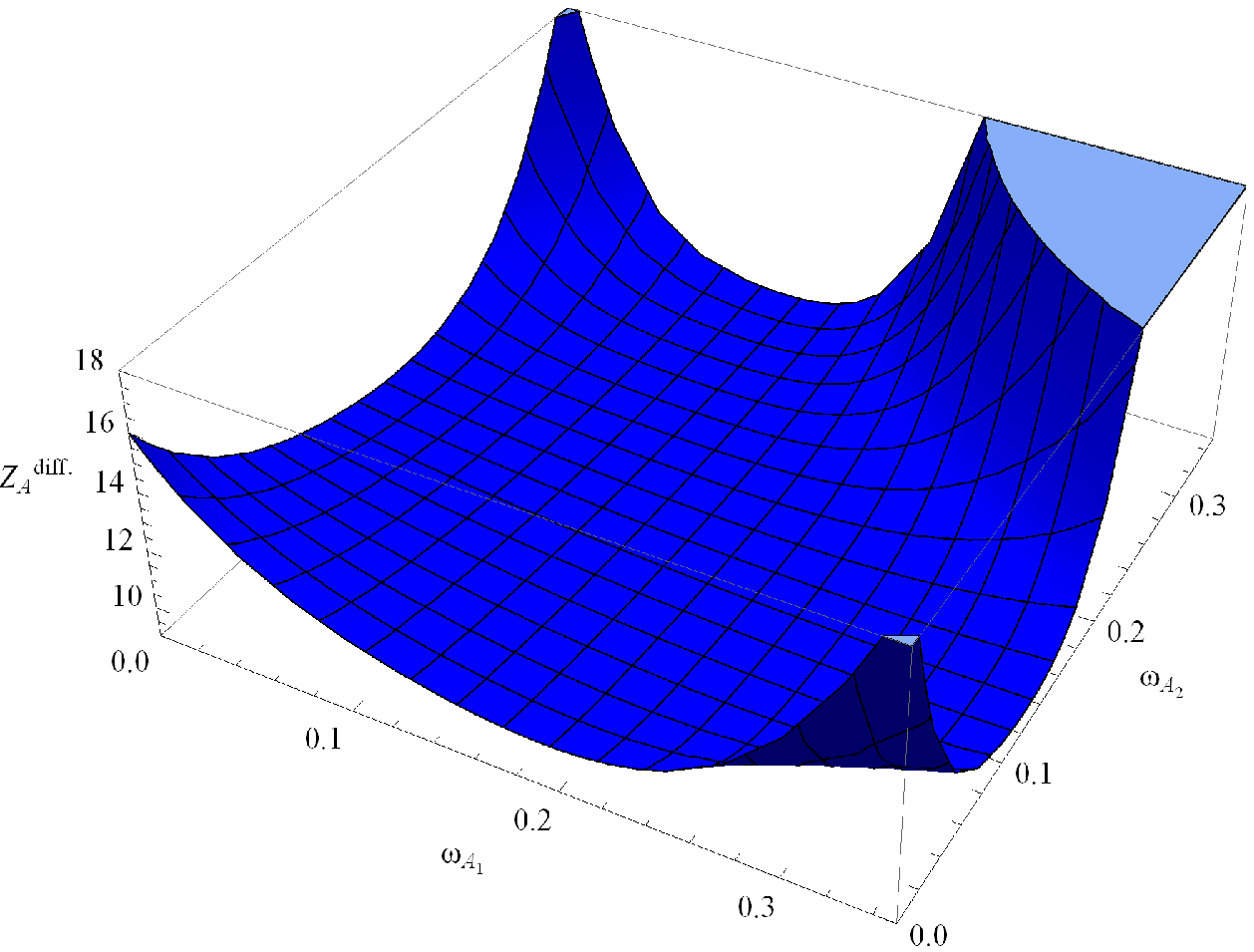,scale=0.5}\\[2mm]
\epsfig{figure=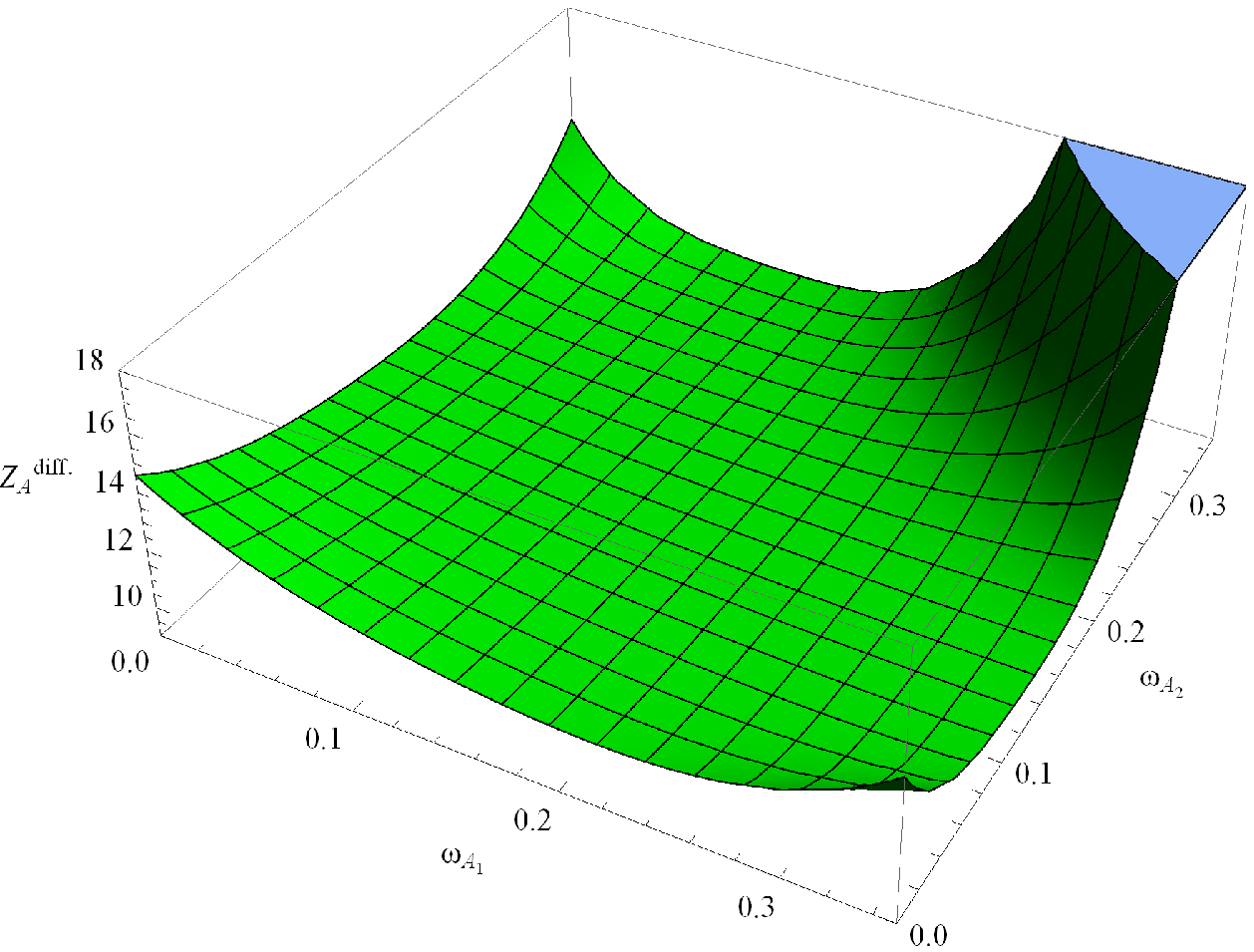,scale=0.5}
\caption{Plots of $Z_A^{\rm{diff.}} \equiv \left[ Z_A^{\rm{(singlet)}} - Z_A^{\rm{(nonsinglet)}} \right] \left( - \frac{g_o^4}{\left( 4 \pi \right)^4} N_f c_F \right)^{-1}$, as a function of $\omega_{A_1}$ and $\omega_{A_2}$ for $\omega_{O_1} = \omega_{O_2} = 0$ $ \ $ (upper left: Wilson action, upper right: TL Symanzik action, lower: Iwasaki action).}
\label{Stagplots3D1}
\end{center}
\end{figure}

\begin{figure}[!ht]
\vspace*{-0.35cm}
\begin{center}
\epsfig{file=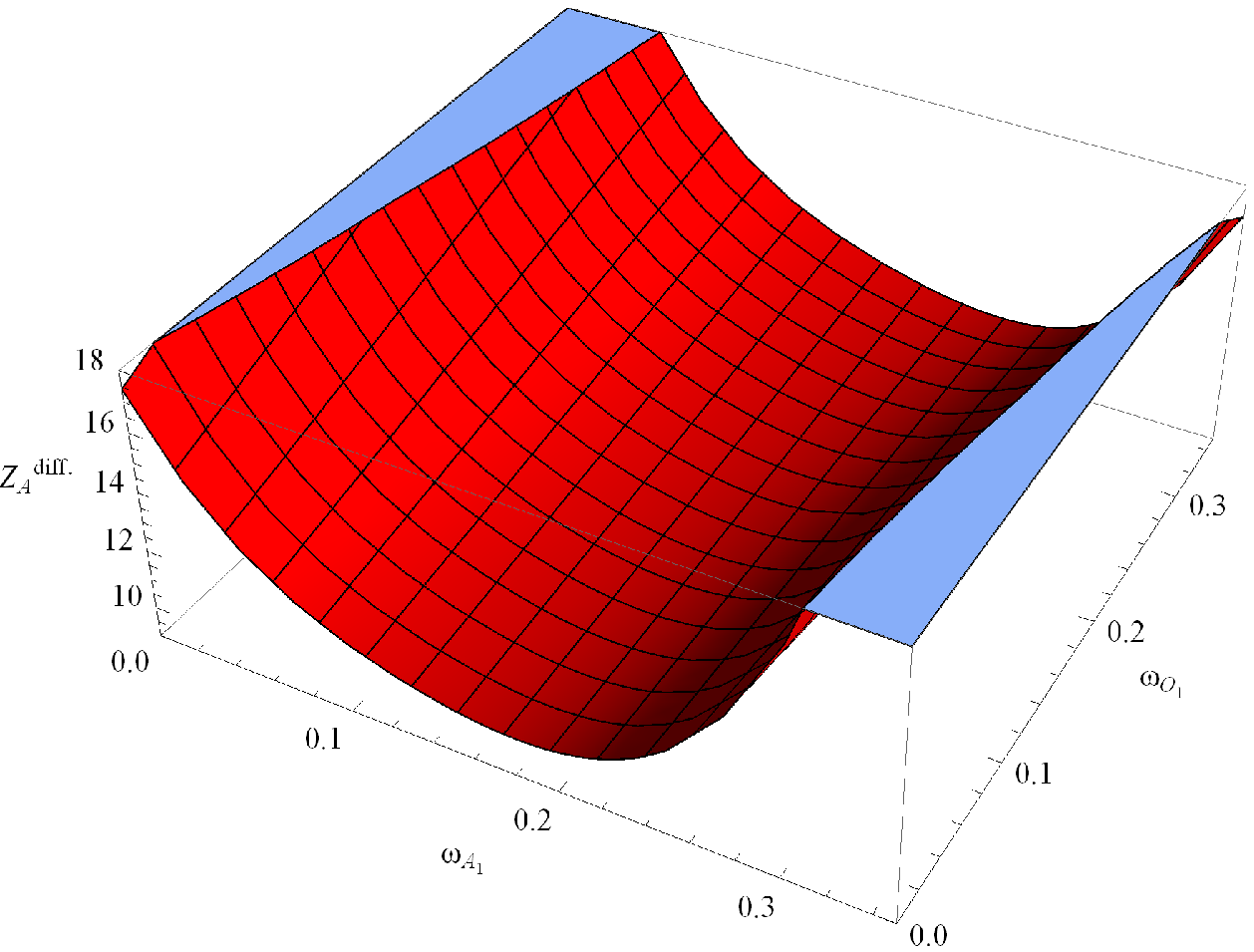,scale=0.5}
\hspace{.05\textwidth}
\epsfig{figure=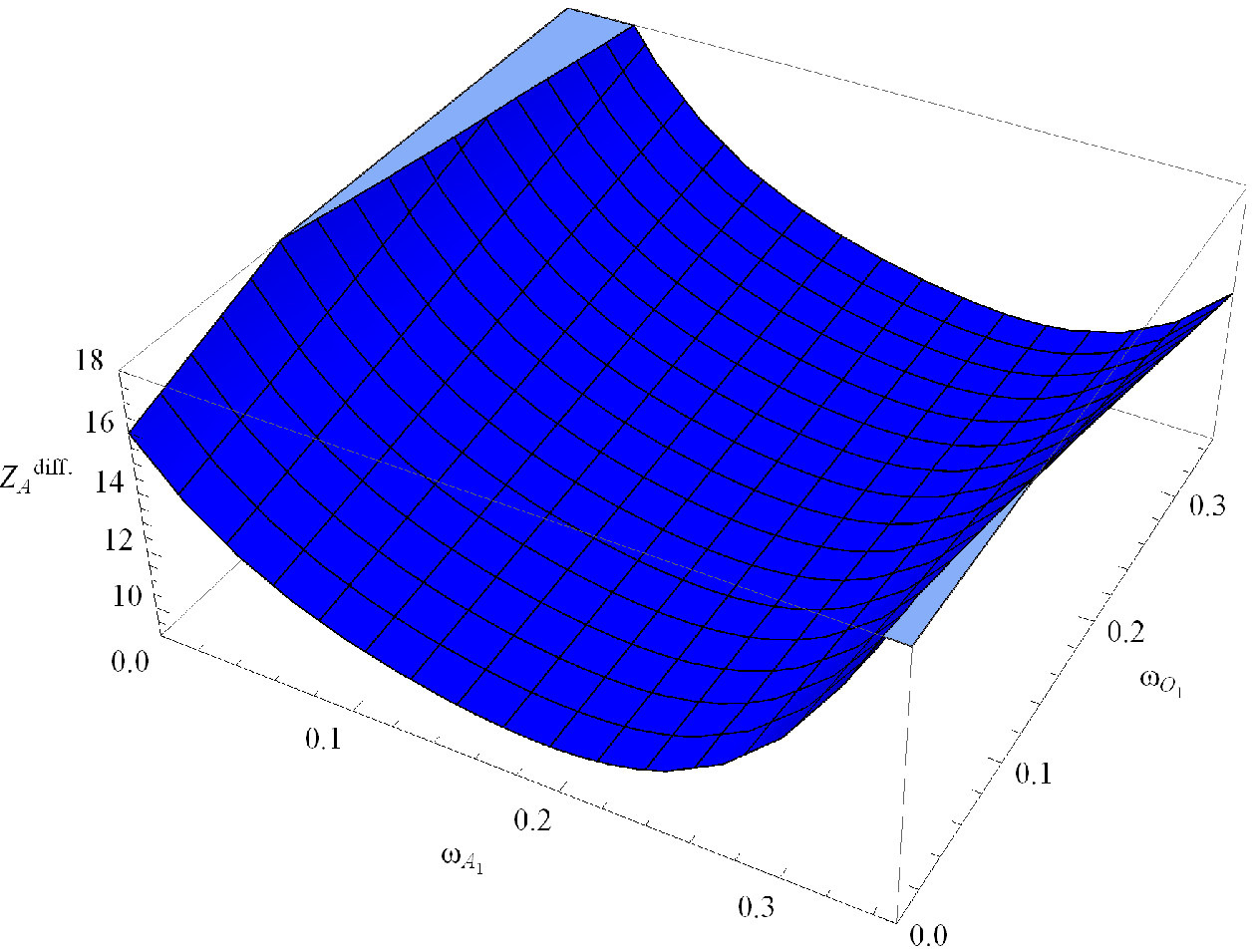,scale=0.5}\\[2mm]
\epsfig{figure=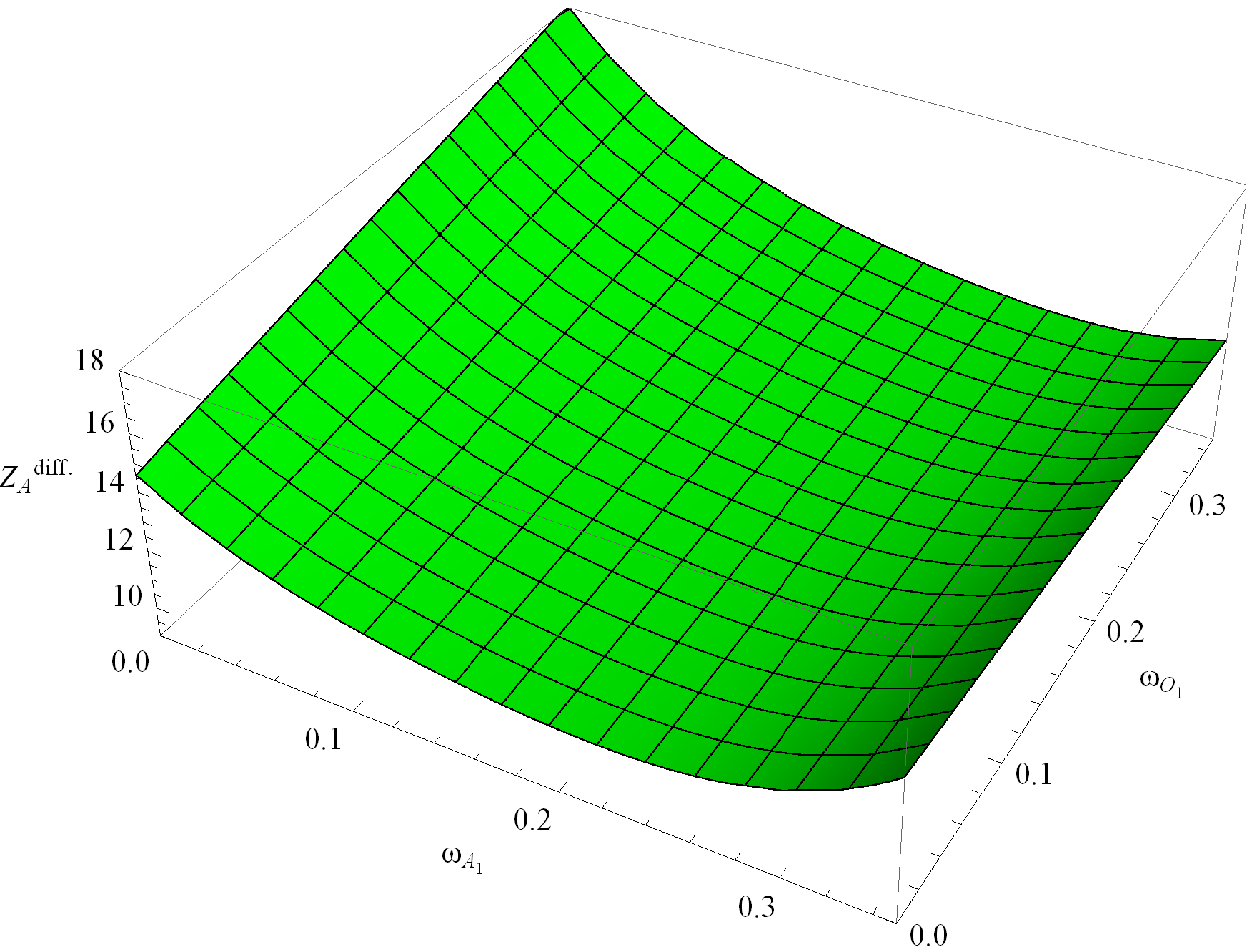,scale=0.5}
\caption{Plots of $Z_A^{\rm{diff.}} \equiv \left[ Z_A^{\rm{(singlet)}} - Z_A^{\rm{(nonsinglet)}} \right] \left( - \frac{g_o^4}{\left( 4 \pi \right)^4} N_f c_F \right)^{-1}$, as a function of $\omega_{A_1}$ and $\omega_{O_1}$ for $\omega_{A_2} = \omega_{O_2} = 0$ $ \ $ (upper left: Wilson action, upper right: TL Symanzik action, lower: Iwasaki action).}
\label{Stagplots3D2}
\end{center}
\end{figure}

\begin{figure}[!ht]
\begin{center}
\epsfig{file=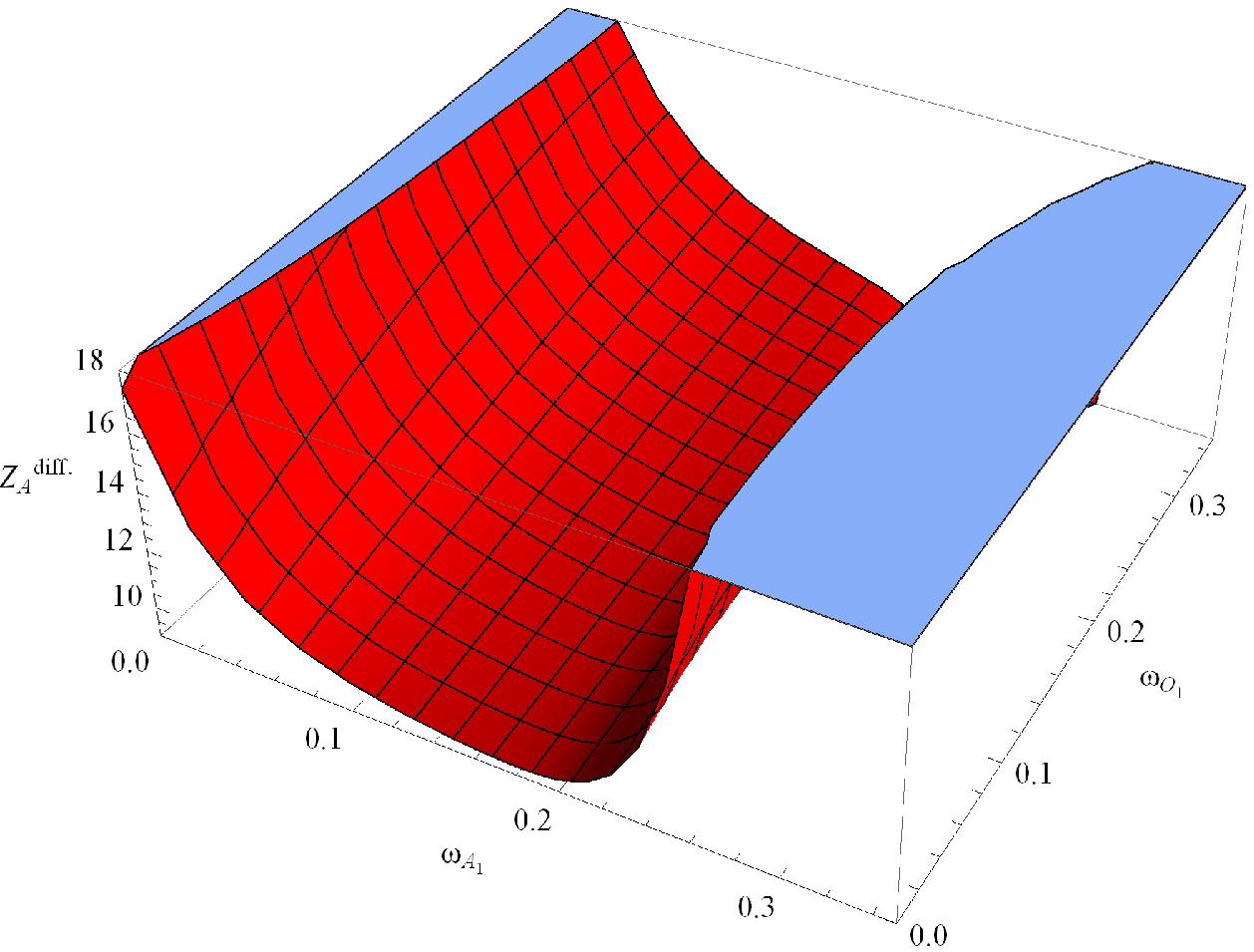,scale=0.5}
\hspace{.05\textwidth}
\epsfig{figure=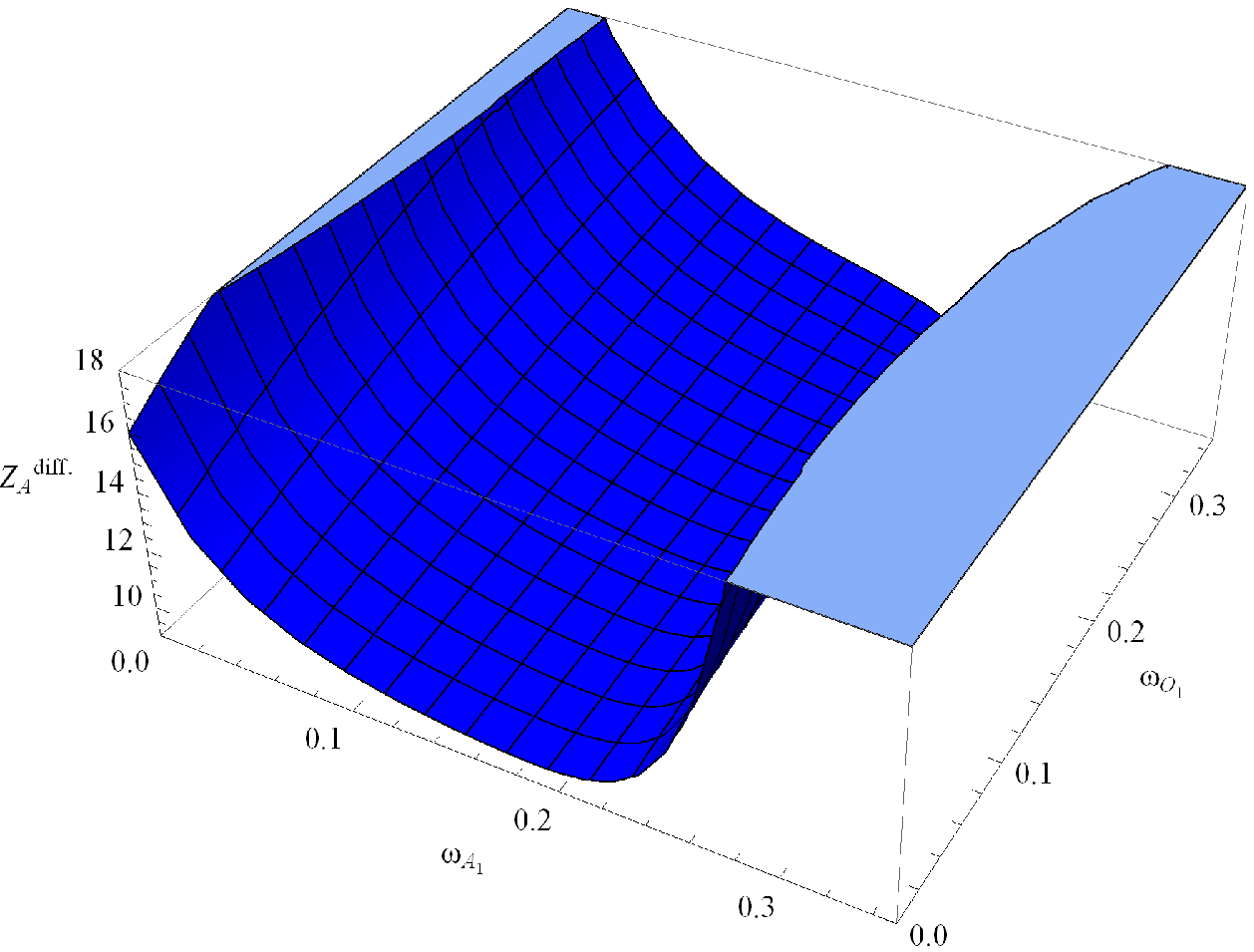,scale=0.5}\\[2mm]
\epsfig{figure=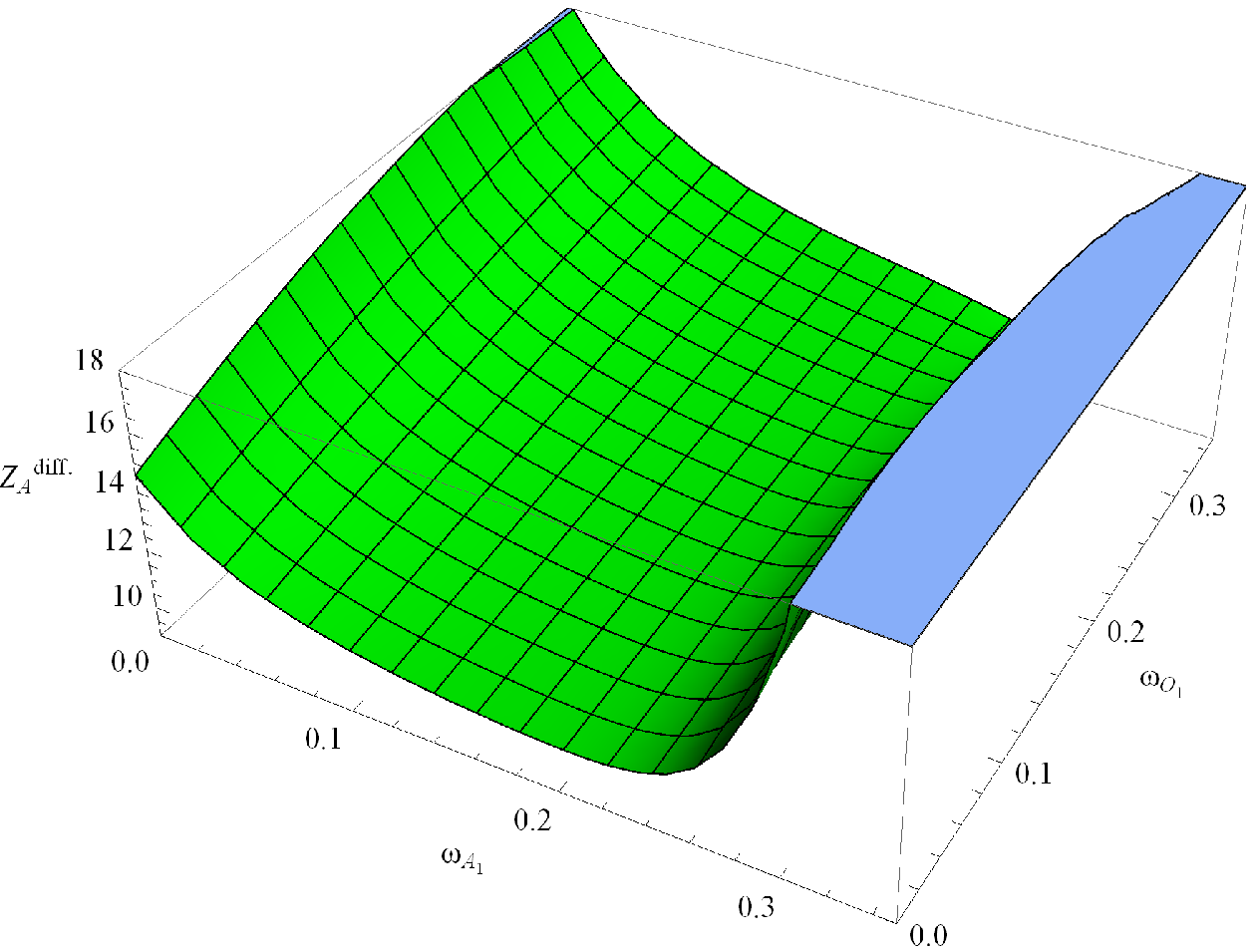,scale=0.5}
\caption{Plots of $Z_A^{\rm{diff.}} \equiv \left[ Z_A^{\rm{(singlet)}} - Z_A^{\rm{(nonsinglet)}} \right] \left( - \frac{g_o^4}{\left( 4 \pi \right)^4} N_f c_F \right)^{-1}$, as a function of $\omega_{A_1}$ and $\omega_{O_1}$ for $\omega_{A_2} = \omega_{A_1}$ and $\omega_{O_2} = \omega_{O_1}$ $ \ $ (upper left: Wilson action, upper right: TL Symanzik action, lower: Iwasaki action).}
\label{Stagplots3D3}
\end{center}
\end{figure}

In Figs. (\ref{Stagplots3D1} - \ref{Stagplots3D3}) we present 3D graphs of our results by selecting
the following parameter values: \\
\hspace*{0.75cm} Fig. \ref{Stagplots3D1}: $\omega_{A_1}, \omega_{A_2}$: free parameters and $\omega_{O_1} = \omega_{O_2} = 0$ (No smearing procedure in the links of operators)\\
\hspace*{0.75cm} Fig. \ref{Stagplots3D2}: $\omega_{A_1}, \omega_{O_1}$: free parameters and $\omega_{A_2} = \omega_{O_2} = 0$ (One smearing step in the links of fermion action and operators)\\
\hspace*{0.75cm} Fig. \ref{Stagplots3D3}: $\omega_{A_1} = \omega_{A_2}, \ \omega_{O_1} = \omega_{O_2}$.\\ 
Just as in 2D graphs, the vertical axis of theses plots corresponds to
$Z_A^{\rm{diff.}}$ for $\bar{\mu} = 1 / a$. We notice again that the plots for the Iwasaki action are flatter than the remaining actions. Also, from the first trio of graphs (Fig. \ref{Stagplots3D1}), we notice that there is only one minimum, on the $45^{\circ}$ axis. Therefore, the two smearing steps of the fermion action give better results than only one smearing step. Also, in Fig. \ref{Stagplots3D2}, as well as in Fig. \ref{Stagplots3D3}, we observe that the stout smearing of the action is more effective in minimizing $Z_A^{\rm{diff.}}$ than the stout smearing of operators. 

\newpage

Further extensions of the present work include the application to
other actions currently used in numerical simulations, including
actions with more steps of stout smearing. In these cases, 
additional contributions to the renormalization functions are more
convergent, and thus their perturbative treatment is simpler;
nevertheless, the sheer size of the vertices (already with 2 stout-smearing
steps we have encountered $\sim 10^6$ terms) 
renders the computation quite cumbersome.
Another possible extension of this work regards 
several variants of staggered fermion action (e.g., HYP smearing \cite{Hasenfratz:2001hp}, 
HEX smearing \cite{Capitani:2006ni}, Asqtad \cite{Orginos:1999cr}). 
Finally, extended versions of $\bar\psi\Gamma\psi$ may be 
studied; in this case the Feynman diagrams of the Fig. \ref{figDiagS} 
will apply also to Wilson fermions and loop integrands will typically 
contain a plethora of new terms, which however will be convergent.

\newpage
\begin{appendix}

\section{Fermion action and fermion bilinear operators in the staggered formulation}

In this appendix we present the derivation of the staggered fermion action and the definition of fermion bilinear operators in the staggered formulation. We start from the naive fermion action:
\be
S_F = a^4\,\sum_{x,\,\mu}\,\bar{
 \psi}(x) \left(\gamma_\mu D_\mu\right)\psi(x) 
+ a^4\,\sum_{x}\,m\,\bar{\psi}(x)\,\psi (x)
\label{naive_action}
\ee
where the covariant derivative $D_\mu$ is defined as follows:
\be
D_\mu \psi(x)=\frac{1}{2a}\Big[U_\mu (x) \psi (x + a \hat{\mu})
- U_\mu^\dagger (x - a \hat{\mu}) \psi (x - a \hat{\mu})\Big]
\ee
The above naive action presents the well-known doubling problem. 
The standard passage to the staggered action entails the
following change of basis:
\bea
\psi(x) = \gamma_x\,\chi(x)&,&\quad 
\bar\psi(x) = \bar\chi(x)\,\gamma_x^\dagger,\nonumber \\[2ex]
 \gamma_x=\gamma_1^{n_1}\,\gamma_2^{n_2}\,\gamma_3^{n_3}\,\gamma_4^{n_4}&,&
\quad x =(a\,n_1,a\,n_2,a\,n_3,a\,n_4),\quad n_i\,\, \epsilon\,\, {\mathbb Z} \label{transf}
\eea
Using the equalities
\be
\gamma_\mu\,\gamma_x = \eta_\mu(x)\gamma_{x +a\,\hat{\mu}}\quad {\rm and}\quad
\gamma_x^\dagger\,\gamma_x = \openone\,,\qquad \eta_\mu (x)  =
(-1)^{\sum_{\nu < \mu} n_\nu}
\label{equalities}
\ee
the lattice fermion action takes the form:
\be
S_F = a^4 \sum_{x} \sum_{\mu} \frac{1}{2a} \, \overline{\chi} (x) 
\eta_\mu (x) \Big[ U_\mu (x) \chi (x + a \hat{\mu})
- U_\mu^\dagger (x - a \hat{\mu}) \chi (x - a \hat{\mu}) \Big] 
+ a^4 \sum_{x} m \overline{\chi} (x) \chi (x)
\label{SFaction1}
\ee
The absence of Dirac indices in the action leads to the assigning of a single fermion 
field component to each lattice site. Hence, the action contains only four rather than 
sixteen fermion doublers, which are called ``tastes".

\bigskip

In the staggered formalism a physical fermion field $\psi (x)$ with taste components is defined as a linear combination of the single-component fermion fields $\chi (x)$ that live on the corners of 4-dimensional elementary hypercubes of the lattice. In standard notation:
\be
\psi_{\alpha,t}(y) = \frac{1}{2}\,\sum_C\left(\gamma_C\right)_{\alpha,t}\,\chi(y)_C\,,\quad
\chi(y)_C = \frac{1}{2} \sum_{\alpha,t}\left(\xi_C\right)_{\alpha,t}\,\psi_{\alpha,t}(y)
\label{Stagg_rot}
\ee
where $\chi(y)_C\equiv\chi(y+aC)/4$, $y$ denotes the position of a hypercube inside the lattice ($y_\mu\,\in\,\,2{\mathbb Z}$), C denotes the position of a fermion field component within a specific hypercube ($C_\mu \in \{0,1\}$), $\gamma_C = \gamma_1^{C_1}\,\gamma_2^{C_2}\,\gamma_3^{C_3}\,\gamma_4^{C_4}$ , $\ \xi_C = \xi_1^{C_1}\,\xi_2^{C_2}\,\xi_3^{C_3}\,\xi_4^{C_4}$, $\ \xi_\mu = \gamma_\mu^\ast$
, $\ \alpha$ is a Dirac index and $t$ is a taste index. In terms of fermion fields with taste components one can now define fermion bilinear operators as in Eq. \eqref{O_G}: ${\cal O}_{\Gamma,\xi}=\bar \psi (x)\,\left(\Gamma\otimes\xi\right)\,\psi (x)$. Transforming to the staggered basis via Eq. \eqref{Stagg_rot}, we extract Eqs. (\ref{O_general} - \ref{gamma1}). Using the relations $\gamma_\mu \gamma_C = \eta_\mu (C) \gamma_{C + \hat{\mu}}$ and $\rm{tr}$ $( \gamma_C^\dagger \gamma_D ) = 4 \delta_{C,D}$, we calculate the quantity $\left(\overline{\Gamma\otimes\openone}\right)_{CD}$ for each operator $\Gamma$:
\bea
\frac{1}{4}{\rm Tr}\left[\gamma_C^\dagger\,\openone\,\gamma_D\right]
&=& \delta_{C,D}\,,\nonumber\\
\frac{1}{4}\,{\rm Tr}\left[\gamma^\dagger_C\,\gamma_\mu\,\gamma_D\,\right]&=&
\delta_{C,D+_{_2}\hat{\mu}}\,\,\eta_\mu(D)\,,\nonumber\\
\label{Oper2}
\frac{1}{4}{\rm Tr}\left[\gamma_C^\dagger\,\smn\,\gamma_D\right]
&=&\frac{1}{i}\,\delta_{C,D+_{_2}\hat{\mu}+_{_2}\hat{\nu}}\,\,\eta_\nu(D)\,\,\eta_\mu(D+_{_2}\hat{\nu})\,,\nonumber\\
\frac{1}{4}\,{\rm Tr}\left[\gamma^\dagger_C\,\gamma_5\,\gamma_\mu\,\gamma_D\,\right]&=& \delta_{C,D+_{_2}\hat{\mu}+_{_2}(1,1,1,1)}\,\,\eta_\mu(D)\,\eta_1(D+_{_2}\hat{\mu})\,\eta_2(D+_{_2}\hat{\mu})\,\eta_3(D+_{_2}\hat{\mu})\,\eta_4(D+_{_2}\hat{\mu})\,,\nonumber\\
\frac{1}{4}\,{\rm Tr}\left[\gamma^\dagger_C\,\gamma_5\,\gamma_D\,\right]&=&
\delta_{C,D+_{_2}(1,1,1,1)}\,\,\eta_1(D)\,\eta_2(D)\,\eta_3(D)\,\eta_4(D)
\eea
where $a +_{_2} b \equiv (a+b)$ mod $2$. Now, the operators can be written as in Eqs. (\ref{OS2} - \ref{OP2}), where the Scalar, Vector, Tensor, Axial Vector and Pseudoscalar operators contain the average of products of 0, 1, 2, 3 and 4 gluon links, respectively. For
example, the average entering the tensor operator of Eq.~(\ref{OT2})
is:
\be
U_{D+_{_2}\hat{\mu}+_{_2}\hat{\nu}, D} = \frac{1}{2}\left[\,
U^\dagger_\nu(y+aD+_{_2}a\hat{\mu})\,\,U_\mu^\dagger(y+aD)
+ \{ \mu\leftrightarrow \nu\}
\,\right]
\label{paths}
\ee
(Eq. \ref{paths} is valid when $(D+_{_2}\hat{\mu}+_{_2}\hat{\nu})_i \ge D_i$, $i=1,2,3,4$, and
takes a similar form for all other cases.) 

\bigskip

Another aspect of the staggered formalism is the representation of the action as well as the representation of the bilinears in momentum space. In order to do this, we use some useful relations, such as the following equivalent expression of $\eta_\mu (x)$:  
\be
\eta_\mu(x)=e^{i\pi\bar\mu\,n}\,,\quad x=an\,,\quad \bar\mu=\sum_{\nu=1}^{\mu-1} \hat{\nu}
\ee         
Also, the summation over the position of $\mathcal{O}_\Gamma$, followed by Fourier transformation leads to expressions of the form:
\be
\sum_{y_\mu\,\epsilon\,2{\mathbb Z}} e^{i\,y\cdot k} = \frac{1}{16}(2\pi)^4\sum_C
\delta_{2\pi}^{(4)}\left(k + \pi\,C\right)
\ee
where $\delta_{2\pi}^{(4)}\left(k \right)$ stands for the standard periodic
$\delta$-function with nonvanishing support at $k\,{\rm mod}2\pi=0$.
In addition, the summation over the index D in the definition of $\mathcal{O}_\Gamma$, after Fourier transformation, may give expressions such as:
\be
\sum_D e^{-i \pi (C - E) \cdot D} = 16 \ \delta_{C,E}
\ee
where $E=(E_1,E_2,E_3,E_4) , \ E_\mu \in \lbrace 0,1 \rbrace$.
Furthermore, expressions like $e^{i k (D +_{_2} \hat{\mu}) a}$ (for Vector and similar expressions for all other operators), which arise through Fourier transformations of the fermion and the antifermion fields, can be written in the following useful form:
\be
 e^{i k (D +_{_2} \hat{\mu}) a} = e^{i k D a} [ \cos (k_\mu a) + i e^{i \pi D \cdot \hat{\mu}} \sin (k_\mu a) ]
\ee   
Finally, since contributions to the continuum limit come from the neighbourhood
of each of the 16 poles of the external momenta $q$, at $q_\mu =
(\pi/a) C_\mu$, it is useful to define $q'_\mu$ and $C_\mu$ through
\be
q_\mu = q'_\mu+\frac{\pi}{a}C_\mu \quad (\,{\rm mod}(\frac{2\pi}{a})\,),\quad (C_\mu\,\,\epsilon\,\,\{0,1\})
\label{C1}
\ee
where the ``small" (physical) part $q'$ has each of its components
restricted to one half of the Brillouin zone: $-\pi/(2a) \le q'_\mu
\le \pi/(2a)$. Thus, conservation of external momenta takes the form: 
\be
\delta_{2\pi}^{(4)}(a\,q_1 -a\,q_2 +\pi D) =
\frac{1}{a}\delta^{(4)}(q_1' -q_2')\, \prod_\mu \delta_{C_{1\mu}+_{_2}C_{2\mu}+_{_2}D_\mu,0}
\ee


\newpage
\section{Evaluation of a basis of nontrivial divergent two-loop Feynman diagrams in the staggered formalism}

In this appendix we present the procedure that we used to evaluate nontrivial divergent integrals which appeared in our two-loop computation using staggered fermions. In the Wilson case, the two-loop divergent integrals, which appeared, can be expressed in terms of a basis of standard integrals found in Ref. \cite{Luscher:1995np}, along with manipulations found in Ref. \cite{Skouroupathis:2008mf}.
However, in the staggered case, the divergent integrals are not related to those standard integrals in an obvious way. Some further steps are needed to this end.  

\bigskip

In our computation, there appeared 4 types of nontrivial divergent 2-loop integrals, using staggered fermions; they are listed below: 
\bea
{I_1}_{\mu \nu} &=& \int_{- \pi}^{\pi} \frac{d^4 k}{{(2 \pi)}^4} \frac{\circe[k]_{\mu} \ \circe[k]_{\nu}}{(\widehat{k}^2)^2 \  (\widehat{k + a q})^2} \int_{- \pi}^{\pi} \frac{d^4 p}{{(2 \pi)}^4} \frac{1}{\circe[p^2] \ (\circe[p + k])^2} \\
{I_2}_{\mu \nu} &=& \int_{- \pi}^{\pi} \frac{d^4 k}{{(2 \pi)}^4} \frac{\circe[k]_{\mu} \ \sin(a q_\nu)}{(\widehat{k}^2)^2 \ (\widehat{k + a q})^2} \int_{- \pi}^{\pi} \frac{d^4 p}{{(2 \pi)}^4} \frac{1}{\circe[p^2] \ (\circe[p + k])^2} \\
{I_3}_{\mu \nu \rho \sigma} &=& \int_{- \pi}^{\pi} \frac{d^4 k}{{(2 \pi)}^4} \frac{\circe[k]_{\mu} \ \circe[k]_{\nu}}{(\widehat{k}^2)^2 \  (\widehat{k + a q})^2} \int_{- \pi}^{\pi} \frac{d^4 p}{{(2 \pi)}^4} \frac{\sin(2 p_\rho) \ \sin(2 p_\sigma)}{(\circe[p^2])^2 \ (\circe[p + k])^2} \\
{I_4}_{\mu \nu \rho \sigma} &=& \int_{- \pi}^{\pi} \frac{d^4 k}{{(2 \pi)}^4} \frac{\circe[k]_{\mu} \ \sin(a q_\nu)}{(\widehat{k}^2)^2 \ (\widehat{k + a q})^2} \int_{- \pi}^{\pi} \frac{d^4 p}{{(2 \pi)}^4} \frac{\sin(2 p_\rho) \ \sin(2 p_\sigma)}{(\circe[p^2])^2 \ (\circe[p + k])^2}
\eea
where $\widehat{p}^2 = \sum_{\mu} \widehat{p}^2_\mu$, \ $\widehat{p}_\mu = 2 \sin(p_\mu / 2)$, \ $\circe[p]^2 = \sum_\mu \circe[p]^2_\mu$, \ $\circe[p]_\mu = \sin(p_\mu)$ and q is an external momentum. The crucial point is the presence of expressions like $ \circe[p]^{2}$ or $(\circe[p + k])^{2}$ rather than $\widehat{p}^{2}$ or $(\widehat{p + k})^{2}$ in the denominators of the above integrals. This behaviour comes from the tree-level staggered fermion propagator. Also, the other crucial point is the fact that we cannot manipulate these integrals via subtractions of the form:
\be
\frac{1}{\circe[p]^2} = \frac{1}{\overset{}{\widehat{p}}^2} + \Big( \frac{1}{\circe[p]^2} - \frac{1}{\overset{}{\widehat{p}}^2} \Big)
\ee  
in order to express them in terms of a standard tabulated integral plus additional terms which are more convergent; such a procedure is applicable, e.g., in the case of the Wilson fermion propagator $\Big[1 / \Big(\circe[p]^2 + r^2 (\widehat{p}^2)^2 / 4\Big)\Big]$ or in other less divergent integrals with staggered fermion propagators. The reason for which such a subtraction cannot be applied is the existence of potential IR singularities at all corners of the Brillouin zone (not only at zero momentum), in the staggered fermion propagator. Therefore, such a subtraction will not alleviate the divergent behaviour at the remaining corners of the Brillouin zone. 

\bigskip

For the above integrals we followed a different approach. At first, we perform the substitution $p_{\mu} \rightarrow p_{\mu}' + \pi \ C_\mu$, where $- \pi/2 < p_\mu' < \pi / 2$ and $C_\mu \in \lbrace 0,1 \rbrace$, which is the same substitution that we applied to external momenta. Now the integration region for the innermost integral breaks up into 16 regions with range $[- \pi/2, \pi / 2]$; the contributions from these regions are identical. To restore the initial range $[- \pi, \pi]$, we apply the following change of variables: $p_\mu' \rightarrow p_\mu'' = 2 p_\mu'$. Then we obtain:
\bea
{I_1}_{\mu \nu} &=& 16 \int_{- \pi}^{\pi} \frac{d^4 k}{{(2 \pi)}^4} \frac{\circe[k]_{\mu} \ \circe[k]_{\nu}}{(\widehat{k}^2)^2 \  (\widehat{k + a q})^2} \int_{- \pi}^{\pi} \frac{d^4 p}{{(2 \pi)}^4} \frac{1}{\widehat{p}^2 \ (\widehat{p + 2k})^2} \\
{I_2}_{\mu \nu} &=& 16 \int_{- \pi}^{\pi} \frac{d^4 k}{{(2 \pi)}^4} \frac{\circe[k]_{\mu} \ \sin(a q_\nu)}{(\widehat{k}^2)^2 \ (\widehat{k + a q})^2} \int_{- \pi}^{\pi} \frac{d^4 p}{{(2 \pi)}^4} \frac{1}{\widehat{p}^2 \ (\widehat{p + 2k})^2} \\
{I_3}_{\mu \nu \rho \sigma} &=& 64 \int_{- \pi}^{\pi} \frac{d^4 k}{{(2 \pi)}^4} \frac{\circe[k]_{\mu} \ \circe[k]_{\nu}}{(\widehat{k}^2)^2 \  (\widehat{k + a q})^2} \int_{- \pi}^{\pi} \frac{d^4 p}{{(2 \pi)}^4} \frac{\circe[p]_{\rho} \ \circe[p]_{\sigma}}{(\widehat{p}^2)^2 \  (\widehat{p + 2k})^2} \\
{I_4}_{\mu \nu \rho \sigma} &=& 64 \int_{- \pi}^{\pi} \frac{d^4 k}{{(2 \pi)}^4} \frac{\circe[k]_{\mu} \ \sin(a q_\nu)}{(\widehat{k}^2)^2 \ (\widehat{k + a q})^2} \int_{- \pi}^{\pi} \frac{d^4 p}{{(2 \pi)}^4} \frac{\circe[p]_{\rho} \ \circe[p]_{\sigma}}{(\widehat{p}^2)^2 \  (\widehat{p + 2k})^2}
\eea
where we omit the double prime from p. The above integrals are similar to standard divergent integrals, computed in Ref. \cite{Luscher:1995np}. The only difference is the presence of a factor of 2 in the denominators, i.e. $1/ (\widehat{p + 2 k})^2$. This can be treated via subtraction methods. We define:
\bea
A(k) &=& \int_{- \pi}^{\pi} \frac{d^4 p}{{(2 \pi)}^4} \frac{1}{\widehat{p}^2 \ (\widehat{p + k})^2} \\
A_{\rm as} (k) &\equiv& \frac{1}{(4 \pi)^2}[- \ln(k^2) + 2] + P_2 \\
B_{\rho \sigma} (k) &=& \int_{- \pi}^{\pi} \frac{d^4 p}{{(2 \pi)}^4} \frac{\circe[p]_{\rho} \ \circe[p]_{\sigma}}{(\widehat{p}^2)^2 \  (\widehat{p + k})^2} \\
\widetilde{B}_{\rho \sigma} (2k) &\equiv& \frac{1}{2 (4 \pi)^2} \frac{\circe[k]_{\rho} \ \circe[k]_{\sigma}}{\widehat{k}^2} + \delta_{\rho \sigma} [\frac{1}{4} A(2k) - \frac{1}{32} P_1]
\eea
where the values of the numerical constants $P_1$ and $P_2$ are noted in Ref. \cite{Luscher:1995np}. $A_{\rm as} (k)$ and $\widetilde{B}_{\rho \sigma} (2k)$ are asymptotic values of $A(k)$ and $B_{\rho \sigma} (2k)$, respectively:
\be
A(k) = A_{\rm as} (k) + \mathcal{O} (k^2), \qquad B_{\rho \sigma} (2k) = \widetilde{B}_{\rho \sigma} (2k) +  \mathcal{O} (k^2)
\ee
The first two integrals ${I_1}_{\mu \nu}$ and ${I_2}_{\mu \nu}$ contain the quantity $A(2k)$ and the remaining two integrals ${I_3}_{\mu \nu \rho \sigma}$ and ${I_4}_{\mu \nu \rho \sigma}$ the quantity $B_{\rho \sigma} (2k)$. We apply the following subtractions:
\bea
A(2k) &=& A(k) + [A_{\rm as} (2k) - A_{\rm as} (k)] + [A(2k) - A(k) - A_{\rm as} (2k) + A_{\rm as} (k)] \\
B_{\rho \sigma} (2k) &=& \widetilde{B}_{\rho \sigma} (2k) + [B_{\rho \sigma} (2k) - \widetilde{B}_{\rho \sigma} (2k)]
\eea
Integrals ${I_1}_{\mu \nu}$ and ${I_2}_{\mu \nu}$ separate into 3 sub-integrals. The first sub-integral with the quantity $A(k)$ is already computed in Ref. \cite{Luscher:1995np} (for ${I_1}_{\mu \nu}$) or can be converted into standard divergent integrals of Ref. \cite{Luscher:1995np} using integration by parts (for ${I_2}_{\mu \nu}$). The second sub-integral with the quantity $[A_{\rm as} (2k) - A_{\rm as} (k)] =  - \ln 4 / (4 \pi)^2$ is a one loop divergent integral computed in Ref. \cite{Luscher:1995np} or \cite{Chetyrkin:1981}. The third sub-integral with the quantity $[A(2k) - A(k) - A_{\rm as} (2k) + A_{\rm as} (k)] = \mathcal{O} (k^2)$ is convergent and so we can integrate it numerically for $a \rightarrow 0$ (In particular, it gives zero for ${I_2}_{\mu \nu}$). Also, integrals ${I_3}_{\mu \nu \rho \sigma}$ and ${I_4}_{\mu \nu \rho \sigma}$ separate into 2 sub-integrals. The first sub-integral with the quantity $\widetilde{B}_{\rho \sigma} (2k)$ gives expressions which can be converted into standard integrals of Refs. \cite{Luscher:1995np, Chetyrkin:1981, Panagopoulos:1990, Ellis:1984} or into the above ${I_1}_{\mu \nu},\ {I_2}_{\mu \nu}$ integrals. The second sub-integral with the quantity $[B_{\rho \sigma} (2k) - \widetilde{B}_{\rho \sigma} (2k)] = \mathcal{O} (k^2)$ is convergent and so we can integrate it numerically for $a \rightarrow 0$ (In particular, it gives zero for ${I_4}_{\mu \nu \rho \sigma}$). Therefore, according to the above manipulations, the final expressions for the four integrals are given by:
\bea
{I_1}_{\mu \nu} &=& \Big\lbrace \frac{2}{(2 \pi)^4} \Big[ - \ln (a^2 q^2) + \frac{3}{2} - \ln 4 \Big] + \frac{1}{2 \pi^2} P_2 \Big\rbrace \frac{q_\mu q_\nu}{q^2} \nonumber \\
&+& \delta_{\mu \nu} \Big\lbrace \frac{2}{(4 \pi)^4} \Big[ \ln(a^2 q^2) \Big]^2 - \frac{1}{4 \pi^2} \Big[P_2 + \frac{1}{(4 \pi)^2} \Big( \frac{5}{2} - \ln 4 \Big) \Big] \ln (a^2 q^2) - \frac{1}{4 \pi^2} \Big[ P_2 + \frac{3}{2 (4 \pi)^2} \ln 4 \Big] + 4 X_2 + G_1 \Big\rbrace \nonumber \\
&+& \mathcal{O} ( a^2 q^2) \\
\nonumber \\
{I_2}_{\mu \nu} &=& \Big\lbrace \frac{1}{(2 \pi)^4} \Big[ \ln (a^2 q^2) - 2 + \ln 4 \Big] - \frac{1}{\pi^2} P_2 \Big\rbrace \frac{q_\mu q_\nu}{q^2} + \mathcal{O} ( a^2 q^2) \\
\nonumber \\
{I_3}_{\mu \nu \rho \sigma} &=& \frac{1}{3 (2 \pi)^4} \frac{q_\mu q_\nu q_\rho q_\sigma}{q^4} + \delta_{\rho \sigma} \Big\lbrace \frac{2}{(2 \pi)^4} \Big[ - \ln (a^2 q^2) + \frac{5}{3} - \ln 4 \Big] - \frac{1}{(4 \pi)^2} (P_1 - 8 P_2) \Big\rbrace \frac{q_\mu q_\nu}{q^2} \nonumber \\
&+& \frac{1}{12 (2 \pi)^4} \Big\lbrace \delta_{\mu \nu} \frac{q_\rho q_\sigma}{q^2} + \delta_{\mu \rho} \frac{q_\nu q_\sigma}{q^2} + \delta_{\mu \sigma} \frac{q_\nu q_\rho}{q^2} + \delta_{\nu \rho} \frac{q_\mu q_\sigma}{q^2} + \delta_{\nu \sigma} \frac{q_\mu q_\rho}{q^2} \Big\rbrace \nonumber \\
&+& \delta_{\mu \nu} \delta_{\rho \sigma} \Big\lbrace \frac{2}{(4 \pi)^4} \Big[ \ln(a^2 q^2) \Big]^2 - \frac{1}{4 \pi^2} \Big[P_2 - \frac{1}{8} P_1 + \frac{1}{(4 \pi)^2} \Big( \frac{51}{2} - \ln 4 \Big) \Big] \ln (a^2 q^2) \nonumber \\
& & \qquad \quad - \frac{1}{4 \pi^2} \Big[ \Big( \frac{1}{3} - \ln 4 \Big) P_2 - \frac{11}{144} P_1 + \frac{3}{2 (4 \pi)^2} \Big( \frac{1}{27} - \ln 4 \Big) \Big] - \frac{1}{2} P_1 \ P_2 + 4 X_2 + G_1 + G_3 \Big\rbrace \nonumber \\
&+& (\delta_{\mu \rho} \delta_{\nu \sigma} + \delta_{\mu \sigma} \delta_{\nu \rho}) \Big\lbrace \frac{1}{(12 \pi)^4} \Big[ - \ln (a^2 q^2) + \frac{1}{6}  \Big] + \frac{1}{6 \pi^2} (P_1 + 3 P_2) + G_2 \Big\rbrace \nonumber \\
&+& \delta_{\mu \nu \rho \sigma} \Big\lbrace \frac{1}{(2 \pi)^4} + \frac{1}{2 (4 \pi)^2} - \frac{1}{3 \pi^2} P_1 + G_4 \Big\rbrace + \mathcal{O} ( a^2 q^2) 
\eea
\bea
{I_4}_{\mu \nu \rho \sigma} &=& - \frac{1}{2 (2 \pi)^4} \frac{q_\mu q_\nu q_\rho q_\sigma}{q^4} - \frac{4}{(4 \pi)^4} \Big\lbrace \delta_{\mu \rho} \frac{q_\nu q_\sigma}{q^2} + \delta_{\mu \sigma} \frac{q_\nu q_\rho}{q^2} \Big\rbrace \nonumber \\
&+& \delta_{\rho \sigma} \Big\lbrace \frac{1}{(2 \pi)^4} \Big[ \ln (a^2 q^2) - \frac{9}{4} \Big] - \frac{1}{2 (2 \pi)^2} (P_1 - 8 P_2) \Big\rbrace \frac{q_\mu q_\nu}{q^2} + \mathcal{O} ( a^2 q^2)
\eea
where $P_1, P_2, X_2$ are given in Ref. \cite{Luscher:1995np} and $G_1 - G_4$ are given below:
\bea
G_1 &=& \phantom{-}0.000803016(6) \\
G_2 &=& -0.0006855532(7) \\
G_3 &=& \phantom{-}0.00098640(7) \\
G_4 &=& \phantom{-}0.00150252(2)
\eea 

\end{appendix}


\bibliographystyle{apsrev}                     
\bibliography{Difference_Z_S-Z_NS}

\begin{thebibliography}{23}
\expandafter\ifx\csname natexlab\endcsname\relax\def\natexlab#1{#1}\fi
\expandafter\ifx\csname bibnamefont\endcsname\relax
  \def\bibnamefont#1{#1}\fi
\expandafter\ifx\csname bibfnamefont\endcsname\relax
  \def\bibfnamefont#1{#1}\fi
\expandafter\ifx\csname citenamefont\endcsname\relax
  \def\citenamefont#1{#1}\fi
\expandafter\ifx\csname url\endcsname\relax
  \def\url#1{\texttt{#1}}\fi
\expandafter\ifx\csname urlprefix\endcsname\relax\def\urlprefix{URL }\fi
\providecommand{\bibinfo}[2]{#2}
\providecommand{\eprint}[2][]{\url{#2}}

\bibitem[{\citenamefont{Constantinou et~al.}(2014)\citenamefont{Constantinou,
  Hadjiantonis, and Panagopoulos}}]{Constantinou:2014rka}
\bibinfo{author}{\bibfnamefont{M.}~\bibnamefont{Constantinou}},
  \bibinfo{author}{\bibfnamefont{M.}~\bibnamefont{Hadjiantonis}},
  \bibnamefont{and}
  \bibinfo{author}{\bibfnamefont{H.}~\bibnamefont{Panagopoulos}},
  \bibinfo{journal}{PoS} \textbf{\bibinfo{volume}{LATTICE2014}},
  \bibinfo{pages}{298} (\bibinfo{year}{2014}), \eprint{[arXiv:1411.6990]}.

\bibitem[{\citenamefont{Chambers et~al.}(2014)\citenamefont{Chambers, Horsley,
  Nakamura, Perlt, Rakow, Schierholz, Schiller, and Zanotti}}]{Chambers:191759}
\bibinfo{author}{\bibfnamefont{A.~J.} \bibnamefont{Chambers}},
  \bibinfo{author}{\bibfnamefont{R.}~\bibnamefont{Horsley}},
  \bibinfo{author}{\bibfnamefont{Y.}~\bibnamefont{Nakamura}},
  \bibinfo{author}{\bibfnamefont{H.}~\bibnamefont{Perlt}},
  \bibinfo{author}{\bibfnamefont{P.~E.~L.} \bibnamefont{Rakow}},
  \bibinfo{author}{\bibfnamefont{G.}~\bibnamefont{Schierholz}},
  \bibinfo{author}{\bibfnamefont{A.}~\bibnamefont{Schiller}}, \bibnamefont{and}
  \bibinfo{author}{\bibfnamefont{J.~M.} \bibnamefont{Zanotti}}
  (\bibinfo{collaboration}{{QCDSF Collaboration}}) (\bibinfo{year}{2014}),
  \eprint{[arXiv:1410.3078]}.

\bibitem[{\citenamefont{Skouroupathis and
  Panagopoulos}(2007)}]{Skouroupathis:2007jd}
\bibinfo{author}{\bibfnamefont{A.}~\bibnamefont{Skouroupathis}}
  \bibnamefont{and}
  \bibinfo{author}{\bibfnamefont{H.}~\bibnamefont{Panagopoulos}},
  \bibinfo{journal}{Phys. Rev.} \textbf{\bibinfo{volume}{D76}},
  \bibinfo{pages}{094514} (\bibinfo{year}{2007}), \eprint{[arXiv:0707.2906]}.

\bibitem[{\citenamefont{Skouroupathis and
  Panagopoulos}(2009)}]{Skouroupathis:2008mf}
\bibinfo{author}{\bibfnamefont{A.}~\bibnamefont{Skouroupathis}}
  \bibnamefont{and}
  \bibinfo{author}{\bibfnamefont{H.}~\bibnamefont{Panagopoulos}},
  \bibinfo{journal}{Phys. Rev.} \textbf{\bibinfo{volume}{D79}},
  \bibinfo{pages}{094508} (\bibinfo{year}{2009}), \eprint{[arXiv:0811.4264]}.

\bibitem[{\citenamefont{Brambilla et~al.}(2014)\citenamefont{Brambilla,
  Di~Renzo, and Hasegawa}}]{DiRenzo:2014}
\bibinfo{author}{\bibfnamefont{M.}~\bibnamefont{Brambilla}},
  \bibinfo{author}{\bibfnamefont{F.}~\bibnamefont{Di~Renzo}}, \bibnamefont{and}
  \bibinfo{author}{\bibfnamefont{M.}~\bibnamefont{Hasegawa}}
  (\bibinfo{year}{2014}), \eprint{[arXiv:1402.6581]}.

\bibitem[{\citenamefont{Aoki et~al.}(2006)\citenamefont{Aoki, Fodor, Katz, and
  Szabo}}]{Aoki:2005vt}
\bibinfo{author}{\bibfnamefont{Y.}~\bibnamefont{Aoki}},
  \bibinfo{author}{\bibfnamefont{Z.}~\bibnamefont{Fodor}},
  \bibinfo{author}{\bibfnamefont{S.~D.} \bibnamefont{Katz}}, \bibnamefont{and}
  \bibinfo{author}{\bibfnamefont{K.}~\bibnamefont{Szabo}},
  \bibinfo{journal}{JHEP} \textbf{\bibinfo{volume}{0601}}, \bibinfo{pages}{089}
  (\bibinfo{year}{2006}), \eprint{[hep-lat/0510084]}.

\bibitem[{\citenamefont{Bors\'anyi et~al.}(2011)\citenamefont{Bors\'anyi,
  Fodor, Katz, Krieg, Ratti et~al.}}]{Borsanyi:2011bm}
\bibinfo{author}{\bibfnamefont{S.}~\bibnamefont{Bors\'anyi}},
  \bibinfo{author}{\bibfnamefont{Z.}~\bibnamefont{Fodor}},
  \bibinfo{author}{\bibfnamefont{S.}~\bibnamefont{Katz}},
  \bibinfo{author}{\bibfnamefont{S.}~\bibnamefont{Krieg}},
  \bibinfo{author}{\bibfnamefont{C.}~\bibnamefont{Ratti}}, \bibnamefont{et~al.}
  (\bibinfo{collaboration}{Wuppertal-Budapest Collaboration}),
  \bibinfo{journal}{J. Phys.} \textbf{\bibinfo{volume}{G38}},
  \bibinfo{pages}{124060} (\bibinfo{year}{2011}), \eprint{[arXiv:1109.5030]}.

\bibitem[{\citenamefont{Bazavov et~al.}(2012)\citenamefont{Bazavov, Bernard,
  DeTar, Freeman, Gottlieb et~al.}}]{Bazavov:2012zad}
\bibinfo{author}{\bibfnamefont{A.}~\bibnamefont{Bazavov}},
  \bibinfo{author}{\bibfnamefont{C.}~\bibnamefont{Bernard}},
  \bibinfo{author}{\bibfnamefont{C.}~\bibnamefont{DeTar}},
  \bibinfo{author}{\bibfnamefont{W.}~\bibnamefont{Freeman}},
  \bibinfo{author}{\bibfnamefont{S.}~\bibnamefont{Gottlieb}},
  \bibnamefont{et~al.} (\bibinfo{collaboration}{MILC Collaboration})
  (\bibinfo{year}{2012}), \eprint{[arXiv:1212.4768]}.

\bibitem[{\citenamefont{Constantinou et~al.}(2013)\citenamefont{Constantinou,
  Costa, and Panagopoulos}}]{Constantinou:2013pba}
\bibinfo{author}{\bibfnamefont{M.}~\bibnamefont{Constantinou}},
  \bibinfo{author}{\bibfnamefont{M.}~\bibnamefont{Costa}}, \bibnamefont{and}
  \bibinfo{author}{\bibfnamefont{H.}~\bibnamefont{Panagopoulos}},
  \bibinfo{journal}{Phys. Rev.} \textbf{\bibinfo{volume}{D88}},
  \bibinfo{pages}{034504} (\bibinfo{year}{2013}), \eprint{[arXiv:1305.1870]}.

\bibitem[{\citenamefont{Horsley et~al.}(2008)\citenamefont{Horsley, Perlt,
  Rakow, Schierholz, and Schiller}}]{Horsley:2008ap}
\bibinfo{author}{\bibfnamefont{R.}~\bibnamefont{Horsley}},
  \bibinfo{author}{\bibfnamefont{H.}~\bibnamefont{Perlt}},
  \bibinfo{author}{\bibfnamefont{P.~E.~L.} \bibnamefont{Rakow}},
  \bibinfo{author}{\bibfnamefont{G.}~\bibnamefont{Schierholz}},
  \bibnamefont{and} \bibinfo{author}{\bibfnamefont{A.}~\bibnamefont{Schiller}},
  \bibinfo{journal}{Phys. Rev.} \textbf{\bibinfo{volume}{D78}},
  \bibinfo{pages}{054504} (\bibinfo{year}{2008}), \eprint{[arXiv:0807.0345]}.

\bibitem[{\citenamefont{Morningstar and Peardon}(2004)}]{Morningstar:2003gk}
\bibinfo{author}{\bibfnamefont{C.}~\bibnamefont{Morningstar}} \bibnamefont{and}
  \bibinfo{author}{\bibfnamefont{M.~J.} \bibnamefont{Peardon}},
  \bibinfo{journal}{Phys. Rev.} \textbf{\bibinfo{volume}{D69}},
  \bibinfo{pages}{054501} (\bibinfo{year}{2004}), \eprint{[hep-lat/0311018]}.

\bibitem[{\citenamefont{Horsley et~al.}(2004)\citenamefont{Horsley, Perlt,
  Rakow, Schierholz, and Schiller}}]{Horsley:2004mx}
\bibinfo{author}{\bibfnamefont{R.}~\bibnamefont{Horsley}},
  \bibinfo{author}{\bibfnamefont{H.}~\bibnamefont{Perlt}},
  \bibinfo{author}{\bibfnamefont{P.~E.~L.} \bibnamefont{Rakow}},
  \bibinfo{author}{\bibfnamefont{G.}~\bibnamefont{Schierholz}},
  \bibnamefont{and} \bibinfo{author}{\bibfnamefont{A.}~\bibnamefont{Schiller}}
  (\bibinfo{collaboration}{QCDSF}), \bibinfo{journal}{Nucl. Phys.}
  \textbf{\bibinfo{volume}{B693}}, \bibinfo{pages}{3} (\bibinfo{year}{2004}),
  \bibinfo{note}{[Erratum: Nucl. Phys. B713, 601(2005)]},
  \eprint{[hep-lat/0404007]}.

\bibitem[{\citenamefont{Patel and Sharpe}(1993)}]{Patel:1992vu}
\bibinfo{author}{\bibfnamefont{A.}~\bibnamefont{Patel}} \bibnamefont{and}
  \bibinfo{author}{\bibfnamefont{S.~R.} \bibnamefont{Sharpe}},
  \bibinfo{journal}{Nucl. Phys.} \textbf{\bibinfo{volume}{B395}},
  \bibinfo{pages}{701} (\bibinfo{year}{1993}), \eprint{[hep-lat/9210039]}.

\bibitem[{\citenamefont{Larin}(1993)}]{Larin:1993tq}
\bibinfo{author}{\bibfnamefont{S.~A.} \bibnamefont{Larin}},
  \bibinfo{journal}{Phys. Lett.} \textbf{\bibinfo{volume}{B303}},
  \bibinfo{pages}{113} (\bibinfo{year}{1993}), \eprint{[hep-ph/9302240]}.

\bibitem[{\citenamefont{Panagopoulos et~al.}(2006)\citenamefont{Panagopoulos,
  Skouroupathis, and Tsapalis}}]{Panagopoulos:2006ky}
\bibinfo{author}{\bibfnamefont{H.}~\bibnamefont{Panagopoulos}},
  \bibinfo{author}{\bibfnamefont{A.}~\bibnamefont{Skouroupathis}},
  \bibnamefont{and} \bibinfo{author}{\bibfnamefont{A.}~\bibnamefont{Tsapalis}},
  \bibinfo{journal}{Phys. Rev.} \textbf{\bibinfo{volume}{D73}},
  \bibinfo{pages}{054511} (\bibinfo{year}{2006}), \eprint{[hep-lat/0601009]}.

\bibitem[{\citenamefont{Constantinou}(2015)}]{Constantinou:2014tga}
\bibinfo{author}{\bibfnamefont{M.}~\bibnamefont{Constantinou}},
  \bibinfo{journal}{PoS} \textbf{\bibinfo{volume}{LATTICE2014}},
  \bibinfo{pages}{001} (\bibinfo{year}{2015}), \eprint{[arXiv:1411.0078]}.

\bibitem[{\citenamefont{Hasenfratz and Knechtli}(2001)}]{Hasenfratz:2001hp}
\bibinfo{author}{\bibfnamefont{A.}~\bibnamefont{Hasenfratz}} \bibnamefont{and}
  \bibinfo{author}{\bibfnamefont{F.}~\bibnamefont{Knechtli}},
  \bibinfo{journal}{Phys. Rev.} \textbf{\bibinfo{volume}{D64}},
  \bibinfo{pages}{034504} (\bibinfo{year}{2001}), \eprint{[hep-lat/0103029]}.

\bibitem[{\citenamefont{Capitani et~al.}(2006)\citenamefont{Capitani, Durr, and
  Hoelbling}}]{Capitani:2006ni}
\bibinfo{author}{\bibfnamefont{S.}~\bibnamefont{Capitani}},
  \bibinfo{author}{\bibfnamefont{S.}~\bibnamefont{Durr}}, \bibnamefont{and}
  \bibinfo{author}{\bibfnamefont{C.}~\bibnamefont{Hoelbling}},
  \bibinfo{journal}{JHEP} \textbf{\bibinfo{volume}{11}}, \bibinfo{pages}{028}
  (\bibinfo{year}{2006}), \eprint{[hep-lat/0607006]}.

\bibitem[{\citenamefont{Orginos et~al.}(1999)\citenamefont{Orginos, Toussaint,
  and Sugar}}]{Orginos:1999cr}
\bibinfo{author}{\bibfnamefont{K.}~\bibnamefont{Orginos}},
  \bibinfo{author}{\bibfnamefont{D.}~\bibnamefont{Toussaint}},
  \bibnamefont{and} \bibinfo{author}{\bibfnamefont{R.~L.} \bibnamefont{Sugar}}
  (\bibinfo{collaboration}{MILC}), \bibinfo{journal}{Phys. Rev.}
  \textbf{\bibinfo{volume}{D60}}, \bibinfo{pages}{054503}
  (\bibinfo{year}{1999}), \eprint{[hep-lat/9903032]}.

\bibitem[{\citenamefont{L{\"u}scher and Weisz}(1995)}]{Luscher:1995np}
\bibinfo{author}{\bibfnamefont{M.}~\bibnamefont{L{\"u}scher}} \bibnamefont{and}
  \bibinfo{author}{\bibfnamefont{P.}~\bibnamefont{Weisz}},
  \bibinfo{journal}{Nucl. Phys.} \textbf{\bibinfo{volume}{B452}},
  \bibinfo{pages}{234} (\bibinfo{year}{1995}), \eprint{[hep-lat/9505011]}.

\bibitem[{\citenamefont{Chetyrkin and Tkachov}(1981)}]{Chetyrkin:1981}
\bibinfo{author}{\bibfnamefont{K.}~\bibnamefont{Chetyrkin}} \bibnamefont{and}
  \bibinfo{author}{\bibfnamefont{F.}~\bibnamefont{Tkachov}},
  \bibinfo{journal}{Nucl. Phys.} \textbf{\bibinfo{volume}{B192}},
  \bibinfo{pages}{159} (\bibinfo{year}{1981}).

\bibitem[{\citenamefont{Panagopoulos and Vicari}(1990)}]{Panagopoulos:1990}
\bibinfo{author}{\bibfnamefont{H.}~\bibnamefont{Panagopoulos}}
  \bibnamefont{and} \bibinfo{author}{\bibfnamefont{E.}~\bibnamefont{Vicari}},
  \bibinfo{journal}{Nucl. Phys.} \textbf{\bibinfo{volume}{B332}},
  \bibinfo{pages}{261} (\bibinfo{year}{1990}).

\bibitem[{\citenamefont{Ellis and Martinelli}(1984)}]{Ellis:1984}
\bibinfo{author}{\bibfnamefont{R.}~\bibnamefont{Ellis}} \bibnamefont{and}
  \bibinfo{author}{\bibfnamefont{G.}~\bibnamefont{Martinelli}},
  \bibinfo{journal}{Nucl. Phys.} \textbf{\bibinfo{volume}{B235}},
  \bibinfo{pages}{93} (\bibinfo{year}{1984}).

\end{thebibliography}

\setcounter{equation}{35}
\newpage
\section*{ADDENDUM: LIST OF CONVERSION RELATIONS BETWEEN $RI'$, $RI'$-$ALTERNATIVE$ AND $\overline{MS}$ SCHEMES}
The two-loop difference $Z_{\Gamma}^{singlet} - Z_{\Gamma}^{nonsinglet}$ is renormalization scheme independent for all operators $\Gamma$, except for the axial vector case. There are two factors contributing to this scheme dependence: On one hand, the conversion factor $Z_5^A$ between $RI'$ and $\overline{MS}$ differs for the singlet and nonsinglet operators (see Eqs. \ref{ZPA},\ref{Z5As},\ref{Z5Ans}); on the other hand, non-identical contributions of the form $\gamma_5 q_{\mu} \slashed{q} / q^2$ in the Green's functions for the singlet and nonsinglet axial vector operators lead to a nontrivial conversion between the $RI'$ and $RI'$-$alternative$ schemes, as shown in Eq. \eqref{ZAvdiff}. In order to clarify scheme dependence, we list below all relevant conversion factors relating the $RI'$, $RI'$-$alternative$ and $\overline{MS}$ schemes. Given that the conversion factors are regularization independent, the renormalization functions $Z_{\Gamma}$ appearing below may be evaluated in any regularization scheme.

\subsection{Conversion factors between $RI'$ and $\overline{MS}$ schemes}
\be
C_{\Gamma}^{\overline{MS},RI'} \equiv C_{\Gamma} \equiv \frac{Z_{\Gamma}^{RI'}}{Z_{\Gamma}^{\overline{MS}}}
\ee
\bea 
C_{S (singlet)}^{\overline{MS},RI'} = C_{S (nonsinglet)}^{\overline{MS},RI'} = 1 + \frac{g_{RI'}^2}{(4 \pi)^2} \ c_F \ (\alpha_{\overline{MS}} + 4) + \frac{g_{RI'}^4}{24 (4 \pi)^4} c_F \Big[\Big(24 \  \alpha_{\overline{MS}}^2 + 96 \ \alpha_{\overline{MS}} - 288 \ \zeta (3) + 57 \Big) \ c_F  \nonumber \\
+ 166 \ N_f - \Big(18 \ \alpha_{\overline{MS}}^2 + 84 \ \alpha_{\overline{MS}} - 432 \ \zeta (3) + 1285 \Big) \ N_c \Big] + \mathcal{O} (g_{RI'}^6)
\eea
\be
C_{P (singlet/nonsinglet)}^{\overline{MS},RI'} = C_{S (singlet/nonsinglet)}^{\overline{MS},RI'} \ Z_5^P
\ee
\be 
C_{V (singlet)}^{\overline{MS},RI'} = C_{V (nonsinglet)}^{\overline{MS},RI'} = 1 + \mathcal{O} (g_{RI'}^8)
\ee  
\be 
C_{A (singlet/nonsinglet)}^{\overline{MS},RI'} = C_{V (singlet/nonsinglet)}^{\overline{MS},RI'} \ Z_5^{A (singlet/nonsinglet)}
\ee
\bea
C_{T (singlet)}^{\overline{MS},RI'} = C_{T (nonsinglet)}^{\overline{MS},RI'} = 1 + \frac{g_{RI'}^2}{(4 \pi)^2} \ c_F \ \alpha_{\overline{MS}} + \frac{g_{RI'}^4}{216 (4 \pi)^4} \ c_F \ \Big[ \Big(216 \  \alpha_{\overline{MS}}^2 + 4320 \ \zeta (3) - 4815 \Big) \ c_F + 626 \ N_f \nonumber \\
+ \Big(162 \ \alpha_{\overline{MS}}^2 + 756 \ \alpha_{\overline{MS}} - 3024 \ \zeta (3) + 5987 \Big) \ N_c \Big] + \mathcal{O} (g_{RI'}^6) 
\eea
where $Z_5^P$ and $Z_5^{A(singlet/nonsinglet)}$ are given in Eqs. (\ref{Z5P} - \ref{Z5Ans}) and $\zeta (x)$ is Riemann's zeta function. The conversion of gauge parameter $\alpha$ between the two schemes is given by:
\be 
\alpha_{RI'} = \frac{\alpha_{\overline{MS}}}{C_{A_{\mu}}^{\overline{MS},RI'}}
\ee
where the conversion factor $C_{A_{\mu}}^{\overline{MS},RI'}$ for the gluon field $A_\mu$ is given by\footnote{Not to be confused with the conversion factor $C_A^{\overline{MS},RI'}$ for the axial vector operator!}
\be 
C_{A_{\mu}}^{\overline{MS},RI'} \equiv \frac{Z_{A_{\mu}}^{RI'}}{Z_{A_{\mu}}^{\overline{MS}}} = 1 + \frac{g_{RI'}^2}{36 (4 \pi)^2} \Big[ \Big(9 \ \alpha_{\overline{MS}}^2 + 18 \ \alpha_{\overline{MS}} + 97 \Big) \ N_c - 40 \ N_f \Big] + \mathcal{O} (g_{RI'}^4)    
\ee
The renormalized coupling constant in the $RI'$ scheme, $g_{RI'}$\,, is conventionally taken to have the same value as in the ${\overline{MS}}$ scheme, $g_{\overline{MS}}$\,.

\subsection{Conversion factors between $RI'$ and $RI'$-$ALTERNATIVE$ schemes}
\be
C_{\Gamma}^{RI',RI' alter} \equiv \frac{Z_{\Gamma}^{RI' alter}}{Z_{\Gamma}^{RI'}}
\ee
\be
C_{S(singlet)}^{RI',RI' alter} = C_{S(nonsinglet)}^{RI',RI' alter} = C_{P(singlet)}^{RI',RI' alter} = C_{P(nonsinglet)}^{RI',RI' alter} = 1
\ee
\be
C_{V(singlet)}^{RI',RI' alter} = C_{V(nonsinglet)}^{RI',RI' alter} = C_{A(nonsinglet)}^{RI',RI' alter} = 1 - \frac{g_{RI'}^4}{(4 \pi)^4} \ c_F \ \Big( \frac{3}{4} c_F - \frac{251}{36} N_c + \frac{19}{18} N_f \Big) + \mathcal{O} (g_{RI'}^6)
\ee
\be
C_{A(singlet)}^{RI',RI' alter} = 1 - \frac{g_{RI'}^4}{(4 \pi)^4} \ c_F \ \Big( \frac{3}{4} c_F - \frac{251}{36} N_c + \frac{1}{18} N_f \Big) + \mathcal{O} (g_{RI'}^6)
\ee
\be
C_{T(singlet)}^{RI',RI' alter} = C_{T(nonsinglet)}^{RI',RI' alter} = 1 + \mathcal{O} (g_{RI'}^6) 
\ee

\subsection{Conversion between $RI'$-$ALTERNATIVE$ and $\overline{MS}$ schemes}

\be
Z_{\Gamma (singlet/nonsinglet)}^{\overline{MS}} = \frac{Z_{\Gamma (singlet/nonsinglet)}^{RI' alter}}{C_{\Gamma (singlet/nonsinglet)}^{\overline{MS},RI'} \ C_{\Gamma (singlet/nonsinglet)}^{RI',RI' alter}}
\ee

\subsection{Relation of $Z_{\Gamma}^{singlet} - Z_{\Gamma}^{nonsinglet}$ between the $RI'$-$ALTERNATIVE$ and $RI'$ schemes}
\be
Z_{\Gamma}^{RI' alter(singlet)} - Z_{\Gamma}^{RI' alter (nonsinglet)} = Z_{\Gamma}^{RI'(singlet)} - Z_{\Gamma}^{RI' (nonsinglet)} \qquad (\Gamma = S, P)
\ee
\be 
Z_{\Gamma}^{RI' alter(singlet)} - Z_{\Gamma}^{RI' alter (nonsinglet)} = Z_{\Gamma}^{RI'(singlet)} - Z_{\Gamma}^{RI' (nonsinglet)} + \mathcal{O} (g_o^6) \qquad (\Gamma = V, T)
\ee
\be 
Z_{A}^{RI' alter(singlet)} - Z_{A}^{RI' alter (nonsinglet)} = Z_{A}^{RI'(singlet)} - Z_{A}^{RI' (nonsinglet)} + \frac{g_o^4}{(4 \pi)^4} c_F N_f + \mathcal{O} (g_o^6) 
\ee

\subsection{Relation of $Z_{\Gamma}^{singlet} - Z_{\Gamma}^{nonsinglet}$ between the $\overline{MS}$ and $RI'$ schemes}
\be
Z_{\Gamma}^{\overline{MS}(singlet)} - Z_{\Gamma}^{\overline{MS} (nonsinglet)} = Z_{\Gamma}^{RI'(singlet)} - Z_{\Gamma}^{RI' (nonsinglet)} + \mathcal{O} (g_o^6) \qquad (\Gamma = S, P, V, T)
\ee
\be 
Z_{A}^{\overline{MS} (singlet)} - Z_{A}^{\overline{MS} (nonsinglet)} = Z_{A}^{RI'(singlet)} - Z_{A}^{RI' (nonsinglet)} + \frac{g_o^4}{(4 \pi)^4} (- \frac{3}{2} c_F N_f) + \mathcal{O} (g_o^6) 
\ee

\subsection{Relation of $Z_{\Gamma}^{singlet} - Z_{\Gamma}^{nonsinglet}$ between the $\overline{MS}$ and $RI'$-$ALTERNATIVE$ schemes}
\be
Z_{\Gamma}^{\overline{MS}(singlet)} - Z_{\Gamma}^{\overline{MS} (nonsinglet)} = Z_{\Gamma}^{RI' alter (singlet)} - Z_{\Gamma}^{RI' alter (nonsinglet)} + \mathcal{O} (g_o^6) \qquad (\Gamma = S, P, V, T)
\ee
\be 
Z_{A}^{\overline{MS} (singlet)} - Z_{A}^{\overline{MS} (nonsinglet)} = Z_{A}^{RI' alter (singlet)} - Z_{A}^{RI' alter (nonsinglet)} + \frac{g_o^4}{(4 \pi)^4} (- \frac{5}{2} c_F N_f) + \mathcal{O} (g_o^6) 
\ee

\end{document}